\documentclass[aps, prd, showpacs, superscriptaddress, nofootinbib, twocolumn]{revtex4-1}

\usepackage{amsmath,amssymb,amsbsy,amstext,amsfonts,mathrsfs,latexsym}
\usepackage{graphicx}

\begin{document}

\title{Cosmological perturbations and stability of nonsingular cosmologies with limiting curvature}

\author{Daisuke Yoshida}
\email{d.yoshida@physics.mcgill.ca}
\affiliation{Department of Physics, McGill University, Montr\'eal, Qu\'{e}bec, H3A 2T8, Canada}

\author{Jerome Quintin}
\email{jquintin@physics.mcgill.ca}
\thanks{Vanier Canada Graduate Scholar}
\affiliation{Department of Physics, McGill University, Montr\'eal, Qu\'{e}bec, H3A 2T8, Canada}

\author{Masahide Yamaguchi}
\email{gucci@phys.titech.ac.jp}
\affiliation{Department of Physics, Tokyo Institute of Technology, Tokyo 152-8551, Japan}

\author{Robert H. Brandenberger}
\email{rhb@hep.physics.mcgill.ca}
\affiliation{Department of Physics, McGill University, Montr\'eal, Qu\'{e}bec, H3A 2T8, Canada}


\begin{abstract}
We revisit nonsingular cosmologies in which the limiting curvature hypothesis is realized.
We study the cosmological perturbations
of the theory and determine the general criteria for stability.
For the simplest model, we find generic Ostrogradski instabilities unless the action contains the
Weyl tensor squared with the appropriate coefficient.
When considering two specific nonsingular cosmological scenarios
(one inflationary and one genesis model), we find ghost and gradient instabilities
throughout most of the cosmic evolution.
Furthermore, we show that the theory is equivalent to a theory of gravity where the action is a general function
of the Ricci and Gauss-Bonnet scalars, and this type of theory is known to suffer from instabilities
in anisotropic backgrounds.
This leads us to construct a new type of curvature-invariant scalar function.
We show that it does not have Ostrogradski instabilities, and it avoids ghost and gradient instabilities
for most of the interesting background inflationary and genesis trajectories.
We further show that it does not possess additional new degrees of freedom in an anisotropic spacetime.
This opens the door for studying stable alternative nonsingular very early Universe cosmologies.
\end{abstract}

\maketitle

\section{Introduction}

One of the biggest problems with the classical theory of general relativity is the occurrence of
singularities, which are inevitable under realistic assumptions\ \cite{Penrose:1964wq,Hawking:1967ju,Hawking:1973uf}
and which signify the breakdown and the incompleteness of the theory.
The big bang singularity in cosmology and the singularity at the center of a black hole are two well-known instances
of singularities in general relativity that one would like to resolve.
These singularities often find themselves in regions of high density, high energy, and high curvature,
where one may expect the breakdown of the classical theory and the emergence of quantum behavior.
For this reason, there is hope that a quantum theory of gravity would provide the resolution to the otherwise pathological
classical singularities.

Without a proper theory of quantum gravity, one may approach the problem
with an effective theory that could mimic the low-energy behavior of the
full quantum theory. The effective theory could be constructed with one
or more new degrees of freedom, e.g.~a new scalar field. This allows
one to study nonsingular theories of the very early Universe within
effective field theory (EFT)
\cite{Cai:2016thi,Creminelli:2016zwa,Cai:2017tku} as is done in, for
instance, bouncing cosmologies
\cite{Qiu:2011cy,Easson:2011zy,Cai:2012va,Koehn:2015vvy} or genesis scenarios
\cite{Creminelli:2010ba,Creminelli:2012my,Nishi:2014bsa,Pirtskhalava:2014esa,Nishi:2015pta,Kobayashi:2015gga}
with a generalized scalar field such as in Horndeski
\cite{Horndeski:1974wa} and generalized Galileon \cite{Deffayet:2011gz}
theories, whose equivalence was first proved in
\cite{Kobayashi:2011nu}. Alternatively, one may attempt to modify the
Einstein-Hilbert gravity action to include higher-order curvature terms (e.g., \cite{Biswas:2005qr,Biswas:2011ar,Conroy:2016sac}).
Interestingly, this can serve as the basis for implementing the limiting
curvature hypothesis, which seeks to incorporate the idea of a
fundamental limiting length (that would be realized in the full theory
of quantum gravity) into the effective theory for gravity.

In line with special relativity where the speed of light is bounded from above
and quantum mechanics where the Heisenberg uncertainty relation holds,
the idea of the limiting curvature hypothesis comes from the fact that one may expect quantum gravity to possess a fundamental length scale $\ell_f$
below which no measurement can be made and on which all physical observables must be smeared out.
Presumably, this fundamental scale is at least of the order of the Planck length
$\ell_\mathrm{Pl}=\sqrt{\hbar G/c^3}$, although it could be larger.
Taking $\ell_f\sim \ell_\mathrm{Pl}$, it is straightforward to see that
if all curvature invariants are bounded throughout the spacetime manifold
($|R|<\ell_\mathrm{Pl}^{-2}$, $|R_{\mu\nu}R^{\mu\nu}|<\ell_\mathrm{Pl}^{-4}$,
$|\nabla_\rho R_{\mu\nu}\nabla^\rho R^{\mu\nu}|<\ell_\mathrm{Pl}^{-6}$,
$|C_{\mu\nu\rho\sigma}C^{\mu\nu\rho\sigma}|<\ell_\mathrm{Pl}^{-8}$, etc.,
where $R_{\mu\nu}$ is the Ricci tensor, $R$ is the Ricci scalar, $C_{\mu\nu\rho\sigma}$ is the Weyl tensor,
and $\nabla$ is the covariant derivative),
then the spacetime is nonsingular. Indeed, in the well-known cases of the big bang and black hole singularities,
some of the physically measurable curvature invariants such as $R$, $R_{\mu\nu}R^{\mu\nu}$,
and $C^2:= C_{\mu\nu\rho\sigma}C^{\mu\nu\rho\sigma}$ blow up;
hence finding theories in which all invariants are bounded is certainly a necessary condition for constructing nonsingular cosmologies.

Unfortunately, bounding an infinite number of curvature invariants is rather nontrivial. Indeed,
there are well-known instances where low-order curvature invariants are bounded
while higher-order invariants are still unbounded
(e.g., $|R_{\mu\nu}R^{\mu\nu}|<\ell_\mathrm{Pl}^{-4}$ while
$|\nabla_\rho R_{\mu\nu}\nabla^\rho R^{\mu\nu}|\rightarrow\infty$).
This is where the limiting curvature hypothesis comes in handy.
The hypothesis states that\ \cite{Markov82,Markov87,Ginsburg:1988jq,Frolov:1989pf,Frolov:1988vj}
if one finds a theory that allows a finite number of curvature invariants
to be bounded by an explicit construction (e.g., $|R|\leq \ell_\mathrm{Pl}^{-2}$ and $|R_{\mu\nu}R^{\mu\nu}|\leq \ell_\mathrm{Pl}^{-4}$),
and when these invariants take on their limiting values, then any solution of the field equations
reduces to a definite nonsingular solution (e.g., de Sitter space), for which all curvature invariants are automatically bounded.
We note, though, that the assumptions of the limiting curvature hypothesis generally do not ensure that solutions
avoid singularities when curvature scalars are not on their limiting values.

The limiting curvature hypothesis has been used and tested in the context of black hole physics
\cite{Frolov:1988vj,Frolov:1989pf,Morgan:1990yy,Trodden:1993dm,Brandenberger:1995es,Brandenberger:1995hd,Bogojevic:1998ma,Easson:2002tg,Frolov:2016pav,Chamseddine:2016ktu}.
In this context, the geometry outside the black hole horizon is described by the usual Schwarzschild metric,
but inside the event horizon, the black hole singularity is replaced with a nonsingular de Sitter spacetime,
which, in turn, could be the source of a new ``baby'' Friedmann universe.
Similarly, in a cosmological context
\cite{Mukhanov:1991zn,Brandenberger:1993ef,Moessner:1994jm,Brandenberger:1995es,Brandenberger:1995hd,Easson:2006jd},
a nonsingular universe can be constructed, in vacuum,
such that it is asymptotically de Sitter in the past and Minkowski in the future (or vice versa).
This is in line with Penrose's vanishing Weyl tensor conjecture\ \cite{Penrose:1900mp} (see also the discussion in Ref.\ \cite{Hawking:1996jh}),
which states that the Weyl tensor should vanish at the beginning of the universe,
since de Sitter space has $C_{\mu\nu\rho\sigma}C^{\mu\nu\rho\sigma}=0$.
With the addition of matter sources, one can obtain asymptotically de Sitter and Friedmann cosmologies, remaining nonsingular
throughout cosmic time. Recent works also show that the ideas of limiting curvature could allow one to construct nonsingular
bouncing cosmologies\ \cite{Chamseddine:2016uef,Liu:2017puc,Bodendorfer:2017bjt}.

In this paper, we want to revisit nonsingular cosmological models that make use of an effective theory for gravity in which
the limit curvature hypothesis is realized. It was shown in Refs.\ \cite{Mukhanov:1991zn,Brandenberger:1993ef} that interesting
background cosmologies can be found within this framework by constructing a theory in which the curvature invariants
$R$ and $4R_{\mu\nu}R^{\mu\nu}-R^2$ are bounded.
However, these studies did not explore the cosmological perturbations\ \cite{Mukhanov:1990me}
for the action containing the above curvature invariants.
Recent developments in nonsingular cosmology within EFT\ \cite{Cai:2016thi,Creminelli:2016zwa,Cai:2017tku}
have shown that it is often rather difficult to avoid instabilities in the cosmological perturbations
(e.g., see Refs.\ \cite{Libanov:2016kfc,Kobayashi:2016xpl,Akama:2017jsa} for no-go theorems within Galileon and Horndeski theories;
see, also, Refs.\ \cite{Ijjas:2016pad,Ijjas:2016tpn,Ijjas:2016vtq}).
For this reason, one could tend to believe that nonsingular models constructed as in Refs.\ \cite{Mukhanov:1991zn,Brandenberger:1993ef}
are going to be very unstable at the perturbation level, thus rendering the models unviable.

In this work, we will show that the naive models of Refs.\ \cite{Mukhanov:1991zn,Brandenberger:1993ef} are indeed
generically unstable. We will see that minimal extensions in which one also includes the Weyl tensor squared, $C_{\mu\nu\rho\sigma}C^{\mu\nu\rho\sigma}$,
in the curvature invariants are more robust, i.e.~there are fairly large regions of parameter space that are stable.
Yet, there do not seem to exist nonsingular cosmological solutions that remain stable throughout cosmic history,
and moreover, the theory will be shown to be equivalent to a $f(R,\mathcal{G})$ theory of gravity (where $\mathcal{G}$
is the Gauss-Bonnet term), which has unavoidable ghosts\ \cite{DeFelice:2010hg}.
We will then construct a completely new curvature-invariant function and show that it allows for stable nonsingular cosmological
solutions throughout time. There will remain some difficulties though in constructing a physically relevant model
in certain cases.

The outline of this paper is as follows. In Sec.~\ref{sec:models1and2}, we will review the construction of
a nonsingular cosmology in which the limiting curvature hypothesis is realized,
set up the action for the theory, and find the background equations of motion. In particular,
we will discuss two specific scenarios: an inflationary scenario and a genesis scenario.
We will then study the cosmological perturbations, determine the general stability conditions
and check them for specific models. The equivalence with $f(R,\mathcal{G})$ gravity will also be demonstrated.
In Sec.~\ref{sec:model3}, we will construct a new model with a new curvature scalar, derive the resulting cosmological perturbations
and the stability conditions, and comment on the case of an anisotropic background.
We will then discuss the recovery of vacuum Einstein gravity and Friedmann cosmology with
the addition of matter sources. We will end with a summary of the results and a discussion in Sec.~\ref{sec:discussion}.

\section{Nonsingular cosmology with limiting curvature}\label{sec:models1and2}

\subsection{Setup of the theory and background evolution}\label{sec:setupmodel1and2}

The approach taken in Refs.\ \cite{Mukhanov:1991zn,Brandenberger:1993ef} to implement the limiting curvature hypothesis
consists of introducing a finite number of nondynamical scalar fields $\chi_i$, or Lagrange multipliers, such that the action takes the form
\begin{align}
 S=\frac{M_\mathrm{Pl}^2}{2}\int\mathrm{d}^4x~\sqrt{-g}\left[R+\sum_i\chi_iI_i-V(\chi_i)\right]~,
\end{align}
where the $I_i$'s are functions of curvature invariants that can depend on $R$, $R_{\mu\nu}$, $R^\mu{}_{\nu\rho\sigma}$,
$C_{\mu\nu\rho\sigma}$ and combinations and derivatives thereof. Accordingly, given an appropriately chosen potential $V(\chi_i)$,
one can rewrite this action into a general $F(R,R_{\mu\nu}R^{\mu\nu},...)$ effective theory of gravity.
By virtue of the Lagrange multipliers, a given potential imposes constraints on the $I_i$'s,
hence the idea that the right choice of $V(\chi_i)$ can naturally bound the curvature invariants and satisfy the limiting curvature
hypothesis asymptotically. We will give examples where this is realized below.

As is done in Refs.\ \cite{Mukhanov:1991zn,Brandenberger:1993ef}, we are going to consider two nondynamical scalar fields
and start with a general action of the form
\begin{align}
\label{eq:masteraction}
 S=&~\frac{M_\mathrm{Pl}^2}{2}\int\mathrm{d}^4x~\sqrt{-g}\Big[R+\chi_1I_1(\nabla,R^\mu{}_{\nu\rho\sigma})-V_1(\chi_1) \nonumber \\
 &+\chi_2I_2(\nabla,R^\mu{}_{\nu\rho\sigma})-V_2(\chi_2)\Big]+S_\mathrm{m}~,
\end{align}
where $S_\mathrm{m}$ is the action for possible matter sources.
At this point, we do not make any assumption on the functional form of $I_1$ and $I_2$,
but we want them to scale as $R$, so let us require that we recover a certain limit at the background level:
\begin{equation}
\label{eq:Iback}
 I_1^{(0)}=12H^2~,\qquad I_2^{(0)}=-6\dot H~.
\end{equation}
The superscript $(0)$ refers to the metric of a flat\footnote{It is straightforward to generalize this to include curvature
(see Ref.\ \cite{Brandenberger:1993ef}).} Friedmann-Lema\^{i}tre-Robertson-Walker (FLRW) universe,
\begin{equation}
 g_{\mu\nu}^{(0)}\mathrm{d}x^\mu\mathrm{d}x^\nu=-N(t)^2\mathrm{d}t^2+a(t)^2\delta_{ij}\mathrm{d}x^i\mathrm{d}x^j~,
\end{equation}
where $N$ is the lapse function and $a$ is the scale factor.
Accordingly, $H:=\dot a/(Na)$ is the Hubble parameter, and a dot is a derivative with respect to physical time, $t$.
At the background level, we can further ask that $\chi_1=\chi_1^{(0)}(t)$ and $\chi_2=\chi_2^{(0)}(t)$.
The original action then becomes\footnote{We omit the superscript $(0)$ for $\chi_1$ and $\chi_2$ when it is clear that they
represent background quantities.}
\begin{align}
 S^{(0)}=&~\frac{M_\mathrm{Pl}^2}{2}\int\mathrm{d}t\mathrm{d}^3\vec{x}~Na^3\left[12\,(1+\chi_1)\left(\frac{\dot a}{Na}\right)^2\right. \nonumber \\
 &\left.+\,6\,(1-\chi_2)\frac{1}{N}\frac{\mathrm{d}}{\mathrm{d}t}\left(\frac{\dot a}{Na}\right)-V_1-V_2\right]+S_\mathrm{m}^{(0)}~.
\end{align}
Varying $S^{(0)}$ with respect to $\chi_1$ and $\chi_2$ yields the equations of motion (EOMs)
\begin{equation}
 12\left(\frac{\dot a}{a}\right)^2=\frac{\mathrm{d}V_1}{\mathrm{d}\chi_1}
\end{equation}
and
\begin{equation}
 -6\left[\frac{\ddot a}{a}-\left(\frac{\dot a}{a}\right)^2\right]=\frac{\mathrm{d}V_2}{\mathrm{d}\chi_2}~,
\end{equation}
respectively, where we set the lapse function to $N=1$.
Letting $T^\mu{}_{\nu}=\mathrm{diag}(-\varepsilon(t),p(t)\delta^i_{\ j})$,
where $T_{\mu\nu}$ is the stress-energy tensor associated with 
the matter action $S_\mathrm{m}$ and where $\varepsilon$ is the energy density and $p$ is the pressure,
one can then vary the background action with respect to $N$ to find
\begin{equation}
 \frac{\varepsilon}{3M_\mathrm{Pl}^2}=(1-2\chi_1-3\chi_2)\left(\frac{\dot a}{a}\right)^2-\dot\chi_2\left(\frac{\dot a}{a}\right)-\frac{1}{6}(V_1+V_2)~,
\end{equation}
again setting $N=1$. Finally, varying with respect to $a$ gives (setting $N=1$ once more)
\begin{align}
 -\frac{p}{M_\mathrm{Pl}^2}=&~(1-2\chi_1-3\chi_2)\left[2\left(\frac{\ddot a}{a}\right)+\left(\frac{\dot a}{a}\right)^2\right] \nonumber \\
 &-2(2\dot\chi_1+3\dot\chi_2)\left(\frac{\dot a}{a}\right)-\ddot\chi_2-\frac{1}{2}(V_1+V_2)~.
\end{align}
Let us denote $V_1':=\mathrm{d}V_1/\mathrm{d}\chi_1$ and
$V_2':=\mathrm{d}V_2/\mathrm{d}\chi_2$ for shorthand notation from here on.
We can then summarize the set of EOMs as
\begin{align}
 12H^2=&~V_1'(\chi_1)~, \\
 -6\dot H=&~V_2'(\chi_2)~, \\
\label{eq:Friedmann1}
 \frac{\varepsilon}{3M_\mathrm{Pl}^2}=&~(1-2\chi_1-3\chi_2)H^2-\dot\chi_2H \nonumber \\
 &-\frac{1}{6}[V_1(\chi_1)+V_2(\chi_2)]~, \\
\label{eq:Friedmann2}
 -\frac{p}{M_\mathrm{Pl}^2}=&~(1-2\chi_1-3\chi_2)[2\dot H+3H^2]-2(2\dot\chi_1+3\dot\chi_2)H \nonumber \\
 &-\ddot\chi_2-\frac{1}{2}[V_1(\chi_1)+V_2(\chi_2)]~.
\end{align}
In the limit where $\chi_1=\chi_2=V_1=V_2=0$, we note that we recover the usual Friedmann equations, as expected.
Also, in the limit where $\varepsilon=p=0$, one obtains the EOMs in vacuum.
Demanding that $V_1'(\chi_1)>0$ for all $\chi_1$ values so that $H$ is real and looking at an expanding universe (so $H>0$;
this could be generalized to a contracting universe with $H<0$, in which case a minus sign would appear in certain equations),
we can write the EOMs as
\begin{align}
\label{eq:adot}
 \dot a&=a\sqrt{\frac{V_1'}{12}}~,\\
\label{eq:chi1dot}
 \dot\chi_1&=-4\sqrt{\frac{V_1'}{12}}\frac{V_2'}{V_1''}~,\\
\label{eq:chi2dot}
 \dot\chi_2&=-\sqrt{\frac{V_1'}{12}}\left[3\chi_2+2\chi_1-1+\frac{2(V_1+V_2)}{V_1'}+\frac{4\varepsilon}{V_1'M_\mathrm{Pl}^2}\right]~.
\end{align}
Furthermore, the equations can be written in the following form:
\begin{align}
 \frac{\mathrm{d}\chi_2}{\mathrm{d}\chi_1}&=\frac{V_1''}{4V_2'}\left[3\chi_2+2\chi_1-1+\frac{2(V_1+V_2)}{V_1'}+\frac{4\varepsilon}{V_1'M_\mathrm{Pl}^2}\right]~,\\
 \frac{\mathrm{d}\chi_1}{\mathrm{d}a}&=-\frac{4}{a}\frac{V_2'}{V_1''}~,\\
 \frac{\mathrm{d}\chi_2}{\mathrm{d}a}&=-\frac{1}{a}\left[3\chi_2+2\chi_1-1+\frac{2(V_1+V_2)}{V_1'}+\frac{4\varepsilon}{V_1'M_\mathrm{Pl}^2}\right]~.
\end{align}

\subsubsection{Example of inflationary scenario}\label{sec:inflation}

\begin{figure*}
 \centering
 \includegraphics[width=0.45\textwidth]{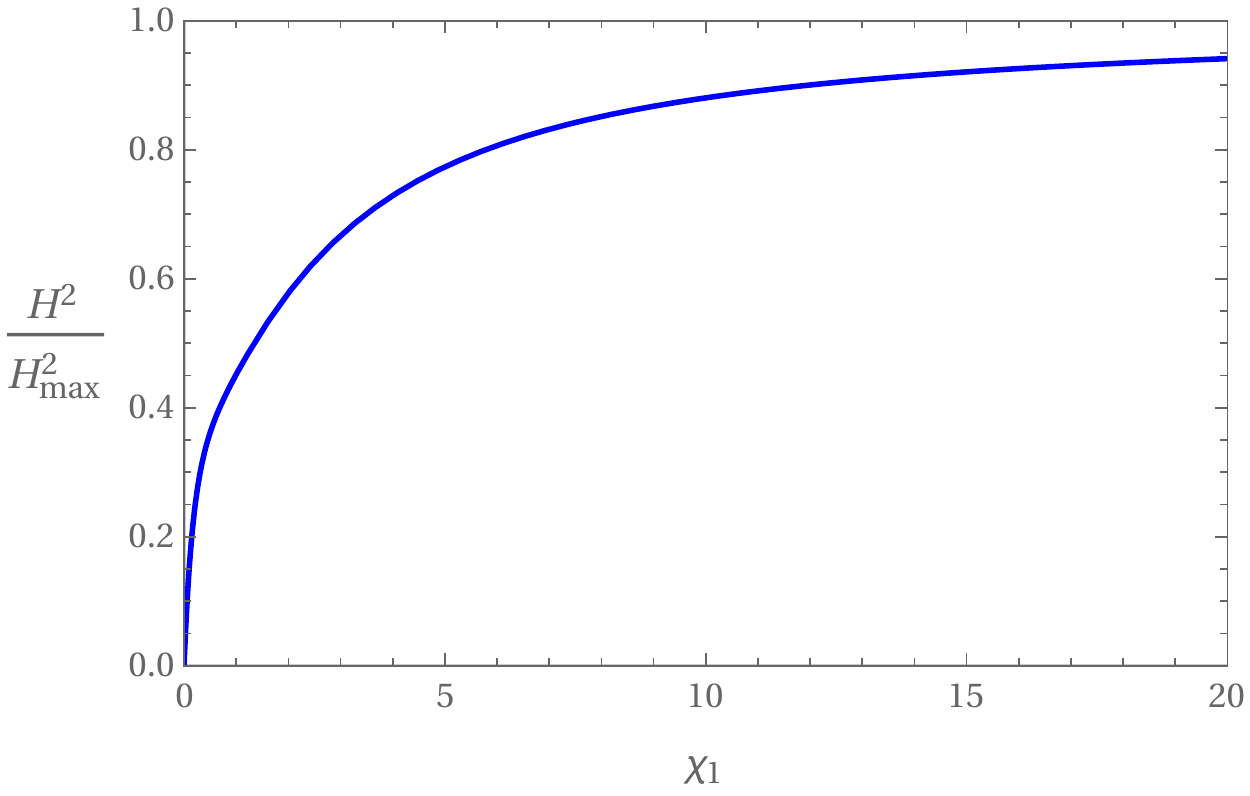}\hfill
 \includegraphics[width=0.45\textwidth]{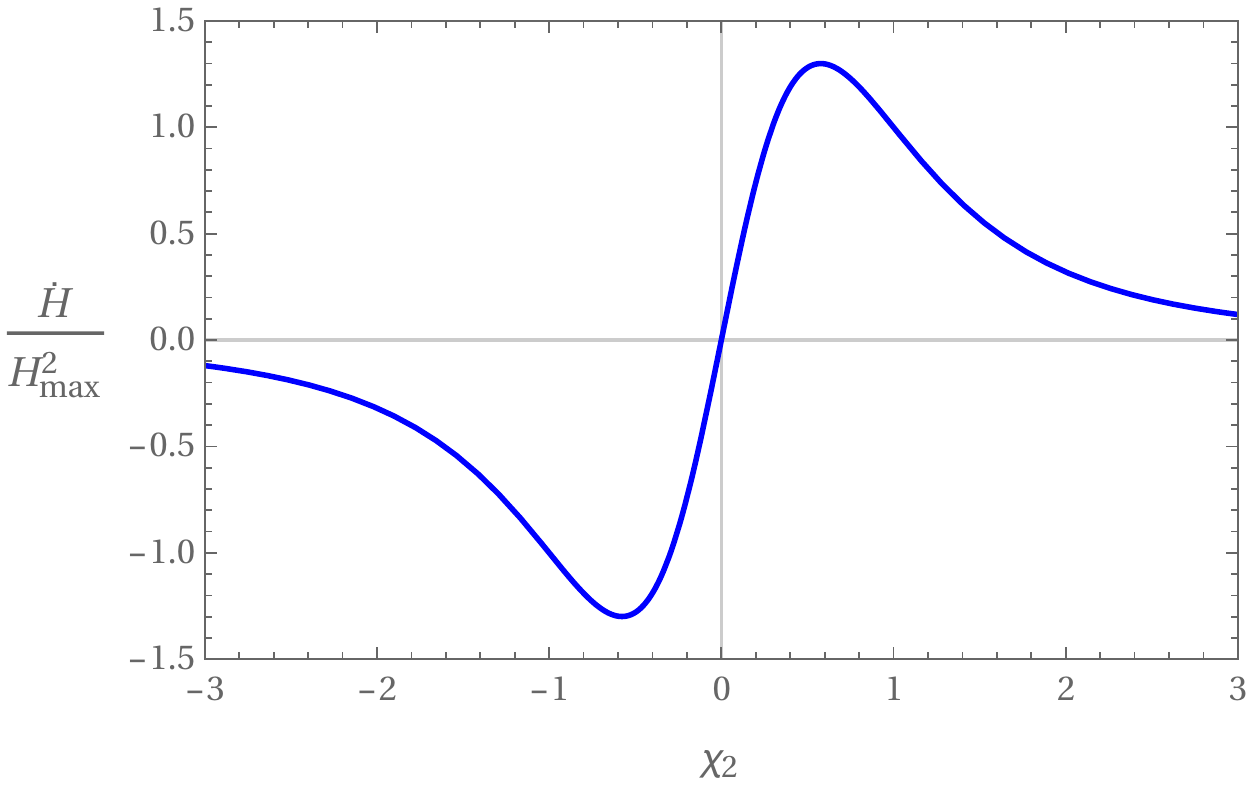}
 \caption{Background trajectories for the inflationary model given by the potentials\ \eqref{eq:V1inf} and\ \eqref{eq:V2inf}.
 The left-hand plot shows $H^2/H_\mathrm{max}^2$ as a function of $\chi_1$ as computed from Eq.\ \eqref{eq:H2inf},
 and the right-hand plot shows $\dot H/H_\mathrm{max}^2$ as a function of $\chi_2$ as computed from Eq.\ \eqref{eq:Hdotinf}.
 Note that $\chi_1$ and $\chi_2$ are dimensionless.}
\label{fig:inflationbackground}
\end{figure*}

To solve the background EOMs, one needs to specify a form for the potentials $V_1(\chi_1)$ and $V_2(\chi_2)$.
As a first example, one can consider
\begin{align}
\label{eq:V1inf}
 V_1(\chi_1)=&~12H_\mathrm{max}^2\frac{\chi_1^2}{1+\chi_1}\left(1-\frac{\ln(1+\chi_1)}{1+\chi_1}\right)~, \\
\label{eq:V2inf}
 V_2(\chi_2)=&-12H_\mathrm{max}^2\frac{\chi_2^2}{1+\chi_2^2}~,
\end{align}
which is inspired from Ref.\ \cite{Brandenberger:1993ef}, and as we will see, this gives rise to a nonsingular inflationary scenario.
In vacuum ($\varepsilon=p=0$), the EOMs are given by
\begin{align}
\label{eq:H2inf}
 \frac{H^2}{H_\mathrm{max}^2}&=\frac{V_1'(\chi_1)}{12H_\mathrm{max}^2}=\frac{2\chi_1+2\chi_1^2+\chi_1^3-2\chi_1\ln(1+\chi_1)}{(1+\chi_1)^3}~, \\
\label{eq:Hdotinf}
 \frac{\dot H}{H_\mathrm{max}^2}&=-\frac{V_2'(\chi_2)}{6H_\mathrm{max}^2}=\frac{4\chi_2}{(1+\chi_2^2)^2}~.
\end{align}
We plot these functions in Fig.~\ref{fig:inflationbackground}.
As we can see in the left-hand plot, the Hubble parameter is finite everywhere: as $\chi_1\rightarrow 0$, the spacetime is asymptotically Minkowski
($H\rightarrow 0$; recall that we are in vacuum), whereas when $\chi_1\rightarrow\infty$, the spacetime is asymptotically de Sitter
since $H\rightarrow H_\mathrm{max}$. Similarly, looking at the right-hand plot, $\dot H$ is finite everywhere,
and it is asymptotically vanishing as $\chi_2\rightarrow\pm\infty$.
Accordingly, this verifies the limiting curvature hypothesis as in Ref.\ \cite{Brandenberger:1993ef}.

We note at this point that since we regard our theory as a low-energy effective theory of a possible quantum theory of gravity,
there should be a cutoff scale beyond which the EFT is no longer valid. The model here includes only two dimensionful parameters:
$M_\mathrm{Pl}$ and $H_\mathrm{max}$. Therefore, the cutoff scale should naively be given by these parameters
as $\Lambda_\mathrm{cut} = (M_\mathrm{Pl} H_\mathrm{max}^n)^{1/(1+n)}$ for a given integer $n\neq -1$,
and in particular, it should be at least of the order of $H_\mathrm{max}$.
Determining the exact value for $\Lambda_\mathrm{cut}$ involves a rather nontrivial computation for the given theory,
but since the energy scale of our cosmological solutions is always less than $H_{\text{max}}$ by construction,
the validity of EFT is naturally ensured.

\begin{figure}
 \centering
 \includegraphics[scale=0.60]{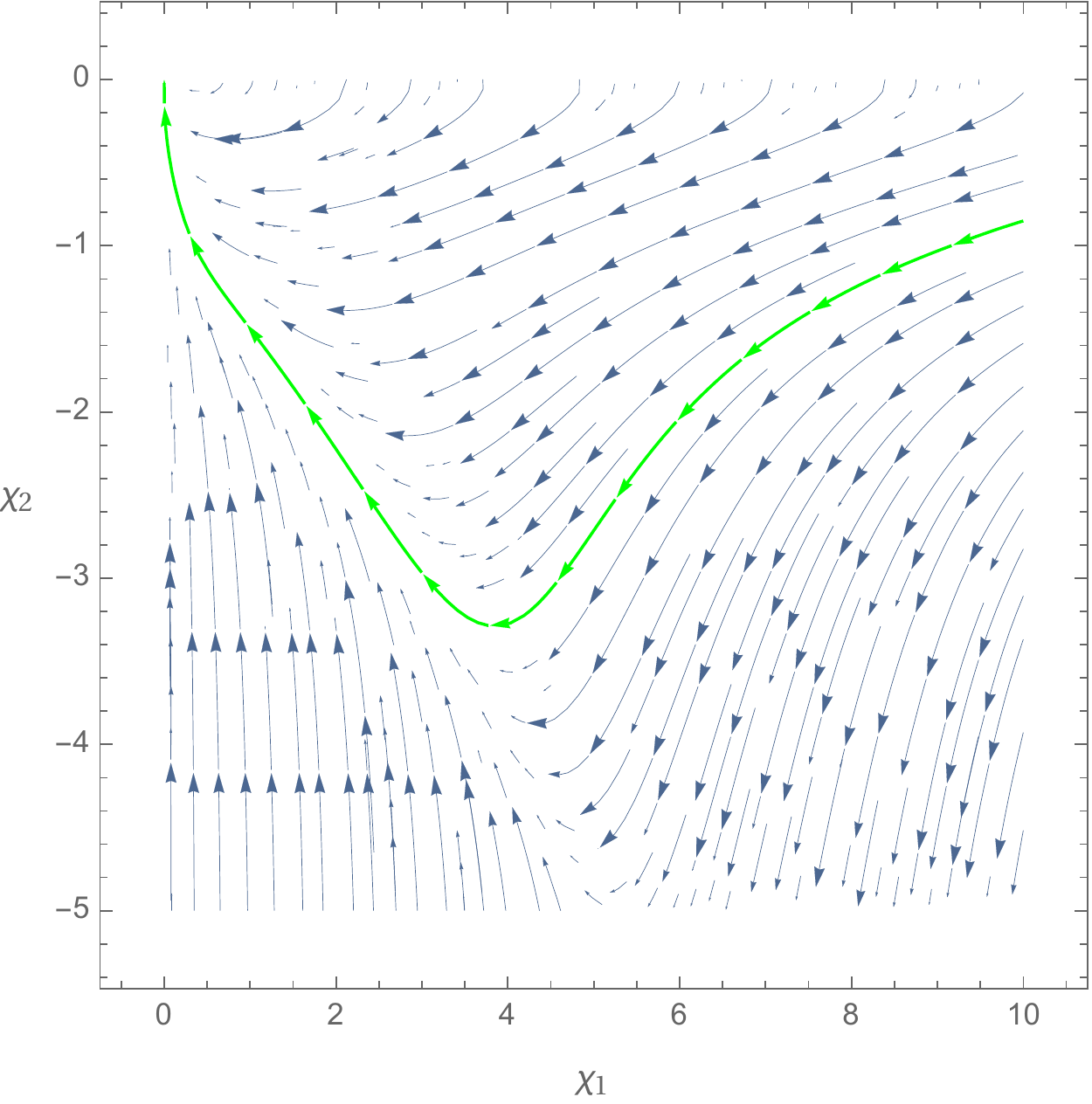}
 \caption{Phase-space diagram of $(\chi_1,\chi_2)$ computed using Eqs.\ \eqref{eq:chi1dot}
 and\ \eqref{eq:chi2dot} showing different background inflationary trajectories.
 The arrows indicate the flow of time. The green curve illustrates a specific trajectory.}
\label{fig:inflationtrajectories}
\end{figure}

The phase-space diagram for the model is plotted in Fig.~\ref{fig:inflationtrajectories},
where the arrows show the vectors with components $(\dot\chi_1,\dot\chi_2)$ computed from Eqs.~\eqref{eq:chi1dot}
and\ \eqref{eq:chi2dot} (in vacuum with $\varepsilon=0$) with the potentials\ \eqref{eq:V1inf} and\ \eqref{eq:V2inf}.
We highlight a specific trajectory in green for illustration. In this case, the universe starts
asymptotically at $\chi_1\rightarrow\infty$ and $\chi_2\rightarrow 0$ and ends asymptotically at
$\chi_1\rightarrow 0$ and $\chi_2\rightarrow 0$, so as we saw from Fig.~\ref{fig:inflationbackground},
the universe starts in a de Sitter spacetime and ends in a Minkowski spacetime.\footnote{With the addition of matter sources,
it would end in a FLRW spacetime as shown in Ref.\ \cite{Brandenberger:1993ef}.}

\begin{figure*}
 \centering
 \includegraphics[width=0.45\textwidth]{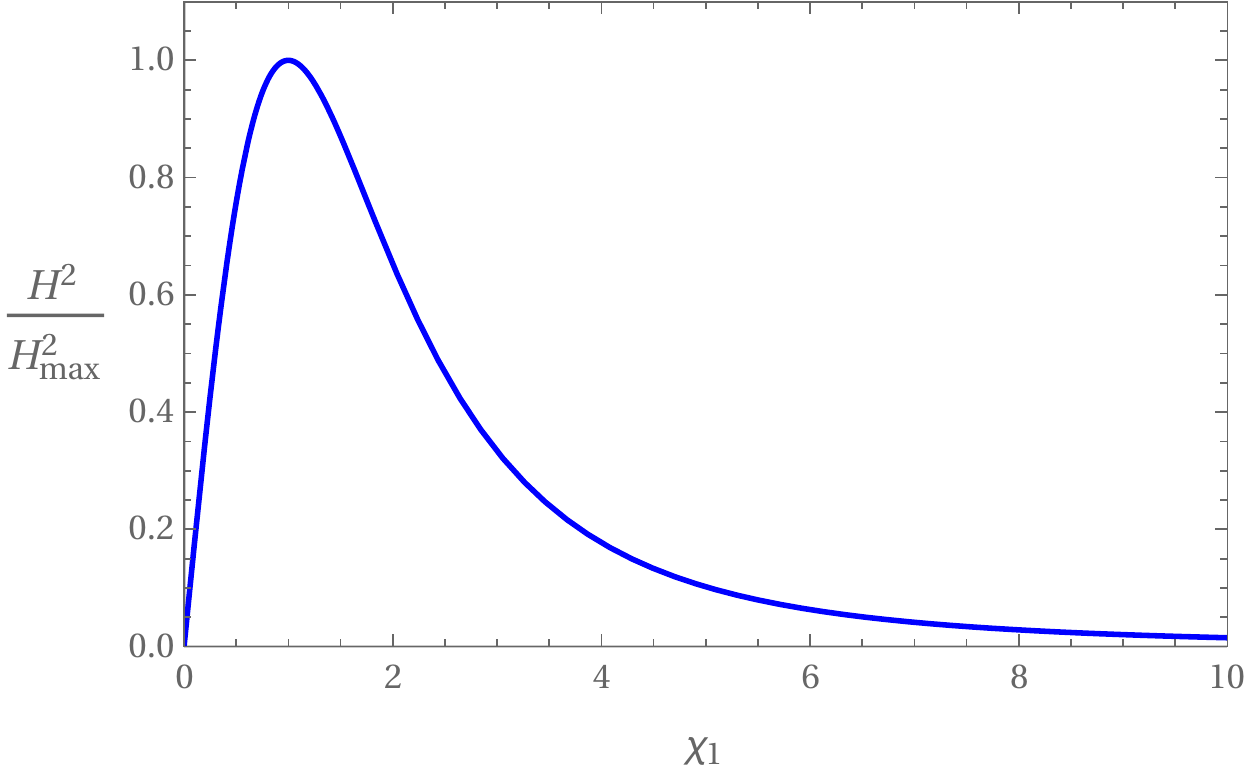}\hfill
 \includegraphics[width=0.45\textwidth]{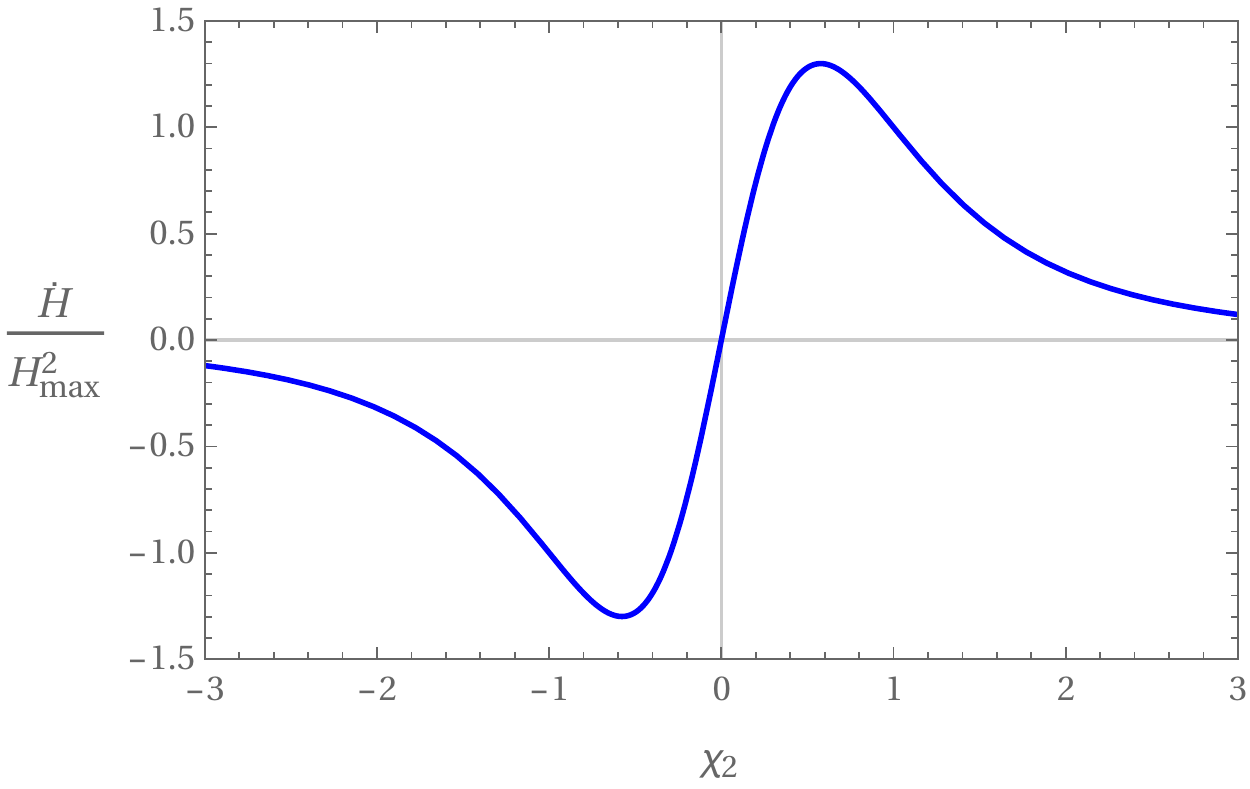}
 \caption{Background trajectories for the genesis model given by the potentials\ \eqref{eq:V1gen} and\ \eqref{eq:V2gen}.
 The left-hand plot shows $H^2/H_\mathrm{max}^2$ as a function of $\chi_1$ as computed from Eq.\ \eqref{eq:H2gen},
 and the right-hand plot shows $\dot H/H_\mathrm{max}^2$ as a function of $\chi_2$ as computed from Eq.\ \eqref{eq:Hdotgen}.}
\label{fig:genesisbackground}
\end{figure*}

At this point, one may wonder how the given scenario evades the singularity theorems
of \cite{Borde:1993xh,Borde:1996pt,Borde:2001nh} regarding the past incompleteness of inflationary cosmology.
First, it is important to recall that the singularity theorem for inflation \cite{Borde:1993xh,Borde:1996pt}
is proved under the assumption that gravity is given by the Einstein-Hilbert action and that inflation is
driven by matter obeying the null energy condition. Our higher derivative gravity terms, when taken to the
matter side of the equations of motion, act as matter violating the null energy condition. Hence the theorem
does not apply in our setup.
We also note that some of the past directed geodesics would have finite affine length (in agreement with
the situation in Ref.\ \cite{Borde:2001nh}), but this is simply due to the fact that
the flat FLRW chart does not cover the entire de Sitter space. One can extend
the spacetime so that all geodesics are complete as in the case of de Sitter space.
Thus, our inflationary universe is free from initial singularities.

\subsubsection{Example of genesis scenario}\label{sec:genesis}

As another example, let us consider
\begin{align}
\label{eq:V1gen}
 V_1(\chi_1)&=-12H_\mathrm{max}^2\frac{8}{3+\chi_1^2}~, \\
\label{eq:V2gen}
 V_2(\chi_2)&=-12H_\mathrm{max}^2\frac{\chi_2^2}{1+\chi_2^2}~.
\end{align}
In vacuum, the EOMs become
\begin{align}
\label{eq:H2gen}
 \frac{H^2}{H_\mathrm{max}^2}&=\frac{V_1'(\chi_1)}{12H_\mathrm{max}^2}=\frac{16\chi_1}{(3+\chi_1^2)^2}~, \\
\label{eq:Hdotgen}
 \frac{\dot H}{H_\mathrm{max}^2}&=-\frac{V_2'(\chi_2)}{6H_\mathrm{max}^2}=\frac{4\chi_2}{(1+\chi_2^2)^2}~.
\end{align}
We plot these functions in Fig.~\ref{fig:genesisbackground}.
As we can see in the left- and right-hand plots, the Hubble parameter and its time derivative are again everywhere finite:
as $\chi_1\rightarrow 0$ or $\chi_1\rightarrow\infty$, the spacetime is asymptotically Minkowski
with $H\rightarrow 0$, and $\dot H\rightarrow 0$ as $\chi_2\rightarrow\pm\infty$.
We note that $H_\mathrm{max}$ is now reached when $\chi_1\rightarrow 1$.
Thus, this is another type of scenario that verifies the limiting curvature hypothesis, namely a genesis scenario,
which starts in Minkowski space rather than de Sitter space.

\begin{figure*}
 \centering
 \includegraphics[width=0.45\textwidth]{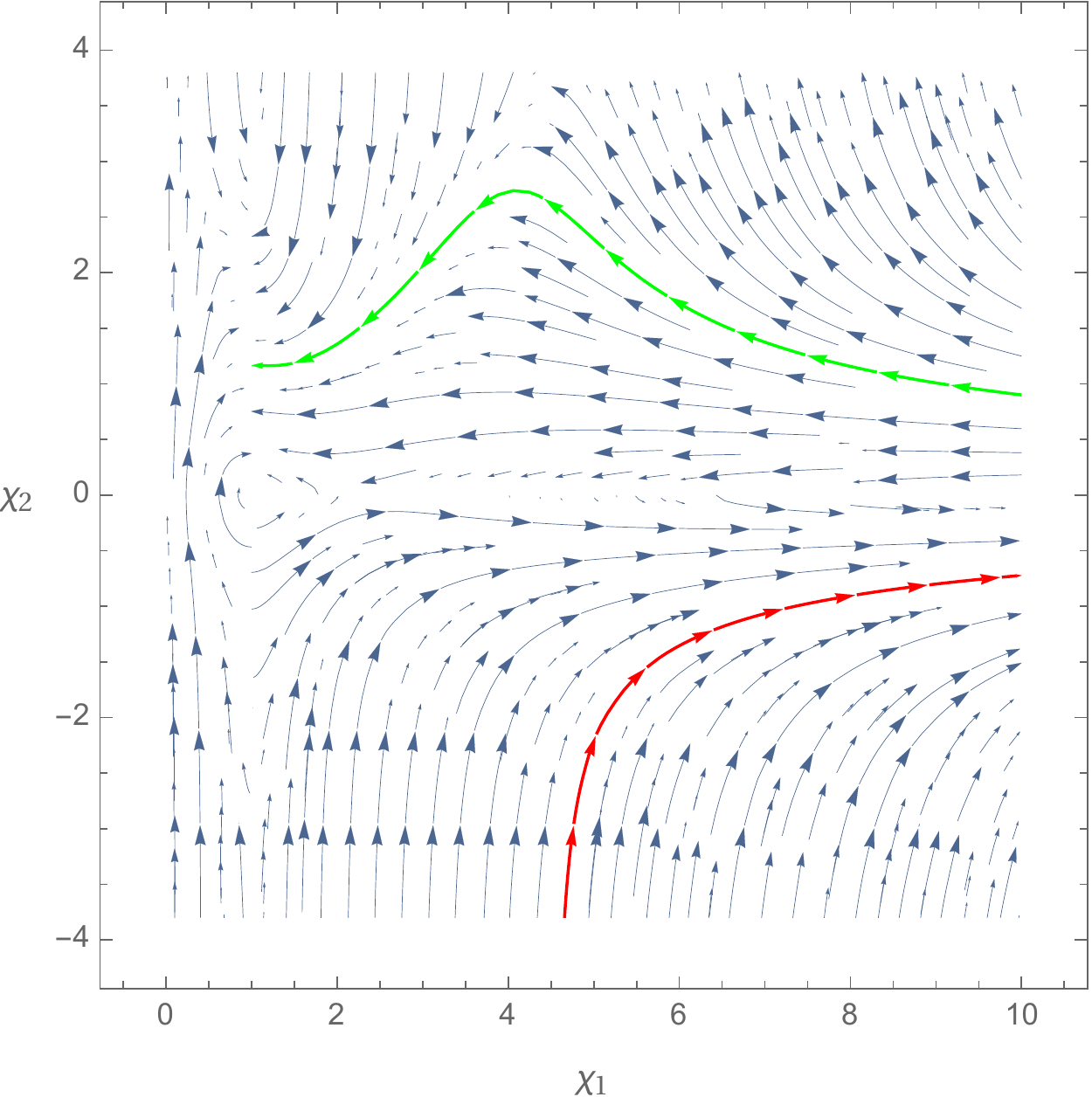}\hfill
 \includegraphics[width=0.45\textwidth]{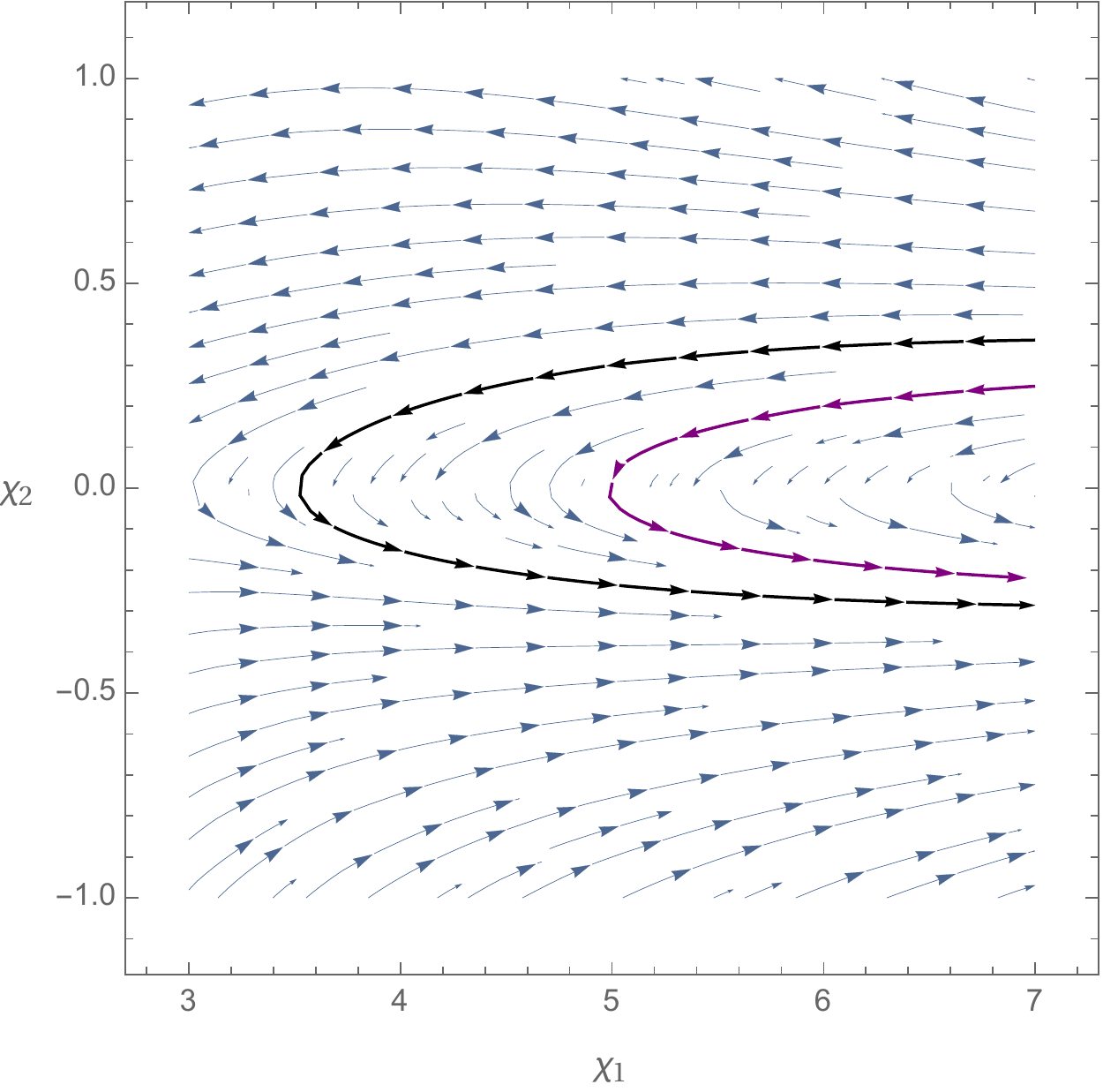}
 \caption{Phase-space diagram of $(\chi_1,\chi_2)$ computed using
 Eqs.\ \eqref{eq:chi1dot},\ \eqref{eq:chi2dot},\ \eqref{eq:V1gen}, and\ \eqref{eq:V2gen}
 showing different background genesis trajectories.
 Note that the right-hand plot is simply a zoomed-in version of the left-hand plot.
 The green, red, black, and purple curves show four different trajectories,
 which are discussed in the text.}
\label{fig:genesistrajectories}
\end{figure*}

The phase-space diagram for this model is plotted in Fig.~\ref{fig:genesistrajectories},
where again, the arrows show the vectors with components $(\dot\chi_1,\dot\chi_2)$ computed from Eqs.~\eqref{eq:chi1dot}
and\ \eqref{eq:chi2dot} (in vacuum with $\varepsilon=0$) with the genesis potentials\ \eqref{eq:V1gen} and\ \eqref{eq:V2gen}.
We highlight different trajectories in green, red, black, and purple for illustration.
All of these curves either start or end at $\chi_1\rightarrow\infty$, which corresponds to Minkowski spacetime.
However, the green and red curves are pathological trajectories since they either end or start at $\chi_1\rightarrow 1$,
at which point it can be shown that $V_1''\rightarrow 0$. Accordingly, from Eq.~\eqref{eq:chi1dot},
one finds $\dot\chi_1\rightarrow\pm\infty$ at that point. More interestingly, the black and purple curves start and end
at $\chi_1\rightarrow\infty$, and they turn around at some minimal $\chi_1>1$ value, so they never reach the ``singularity'' at $\chi_1=1$.
Also, these trajectories always have $\chi_2\ll 1$.
When looking at the left-hand plot of Fig.~\ref{fig:genesisbackground}, these trajectories suggest that the universe starts
in the far right at $\chi_1\rightarrow\infty$ (Minkowski spacetime), rolls up to the left but does not reach $H_\mathrm{max}$,
and rolls back down to Minkowski spacetime again. In light of a structure formation scenario for the very early universe,
one would like to have some form of reheating\footnote{For instance, reheating could occur via
gravitational particle production\ \cite{Parker:1968mv,Parker:1969au} (see Refs.\ \cite{Kobayashi:2015gga,Quintin:2014oea} for examples
of gravitational particle production in nonsingular cosmologies).}
near the maximal value that the Hubble parameter reaches. Thus, the universe would start as Minkowski spacetime, but it would end
as a radiation- and then matter-dominated FLRW spacetime.

\subsection{Cosmological perturbations and stability analysis}\label{sec:stabilitymodels1and2}

We now turn to the study of the cosmological perturbations for the action given by Eq.~\eqref{eq:masteraction}.
At this point, one needs to specify the form of the curvature-invariant functions $I_1$ and $I_2$.
Motivated by Refs.\ \cite{Mukhanov:1991zn,Brandenberger:1993ef}, let us take
\begin{align}
\label{eq:Imodels12}
 I_1&:=R+\sqrt{12R_{\mu\nu}R^{\mu\nu}-3R^2+3\kappa C_{\mu\nu\rho\sigma}C^{\mu\nu\rho\sigma}}~,\nonumber\\
 I_2&:=\sqrt{12R_{\mu\nu}R^{\mu\nu}-3R^2+3\kappa C_{\mu\nu\rho\sigma}C^{\mu\nu\rho\sigma}}~,
\end{align}
where at this point $\kappa$ is just some real constant.
In a flat FLRW background, these curvature invariants reduce to
\begin{equation}
 I_1^{(0)}=12H^2~,\qquad I_2^{(0)}=-6\dot H^2~,
\end{equation}
under the assumption\footnote{We note that the requirement $\dot H<0$ restricts our attention to the regions of phase space in which $\chi_2<0$
for the examples given in Secs.~\ref{sec:inflation} and \ref{sec:genesis}.} that $\dot H<0$,
which was the hypothesis of Eq.~\eqref{eq:Iback}
that allowed us to find the general background EOMs
in Sec.~\ref{sec:setupmodel1and2}.
We note that the background expressions for the curvature invariants
do not depend on the constant $\kappa$ since flat FLRW spacetime is conformally flat,
so the term proportional to the Weyl tensor squared does not affect the dynamics
of background spacetime.

\subsubsection{Tensor modes}

Let us begin by studying the tensor fluctuations. We start by perturbing the metric linearly as
\begin{equation}
\label{eq:deftensor}
 g_{\mu\nu}^{(1)}\mathrm{d}x^\mu\mathrm{d}x^\nu=-\mathrm{d}t^2+a^2\left(\delta_{ij}+h_{ij}+\frac{1}{2}h_{ik}h^k{}_j\right)\mathrm{d}x^i\mathrm{d}x^j~,
\end{equation}
where the perturbation tensor $h_{ij}$ is transverse and traceless, i.e.~$h^i{}_i=\partial_ih^i{}_j=0$ (adding
the last term on the right-hand side does not change the linear equations but simplifies the derivation).
We define the Fourier components of $h_{ij}$ by
\begin{equation}
 h_{ij}=\int\frac{\mathrm{d}^3\vec{k}}{(2\pi)^3}\left[h_{\vec{k}}^{+}e_{ij}^{+}(\vec{k})
 +h_{\vec{k}}^{\times}e_{ij}^{\times}(\vec{k})\right]\mathrm{e}^{\mathrm{i}\vec{k}\cdot\vec{x}}~,
\end{equation}
where $\{e_{ij}^{+},e_{ij}^{\times}\}$ represents the polarization basis.
Given the curvature-invariant functions of Eq.~\eqref{eq:Imodels12},
we can now perturb Eq.~\eqref{eq:masteraction} to second order with the above metric to find
\begin{align}
\label{eq:ST12}
 S_T^{(2)}=&~\frac{M_\mathrm{Pl}^2}{2}\int\mathrm{d}t\frac{\mathrm{d}^3\vec{k}}{(2\pi)^3}\sum_{I=+,\times}a^3
 \Big(\mathcal{G}_T\ddot{h}_{\vec{k}}^I\ddot{h}_{-\vec{k}}^I
 +\mathcal{K}_T\dot{h}_{\vec{k}}^I\dot{h}_{-\vec{k}}^I \nonumber \\
 &-\mathcal{M}_T\frac{k^2}{a^2}h_{\vec{k}}^Ih_{-\vec{k}}^I\Big)~,
\end{align}
where
\begin{equation}
 \mathcal{G}_T=-(2+\kappa)\frac{\chi_1+\chi_2}{4\dot H}~,
\end{equation}
and the coefficients $\mathcal{K}_T$ and $\mathcal{M}_T$ will be specified shortly.
We pause here to note that, in general, since the second-order action in the tensor sector has nondegenerate higher-derivative terms
($\propto\ddot{h}^2$), there appear to be ghost degrees of freedom according to Ostrogradski's theorem.
The only way to avoid those Ostrogradski ghosts would be if $\mathcal{G}_T$ were to vanish identically ($\mathcal{G}_T\equiv 0$).
We see that this is not possible for a generic real constant $\kappa$, but if one sets
$\kappa=-2$, then the model is safe with regards to Ostrogradski instabilities.
The original models of Refs.\ \cite{Mukhanov:1991zn,Brandenberger:1993ef} did not include $C_{\mu\nu\rho\sigma}C^{\mu\nu\rho\sigma}$
in their curvature invariants at all, so they had $\kappa=0$. Accordingly,
the above implies that these models are inherently unstable.
Yet, the addition of the Weyl tensor squared in the invariants with a specific prefactor
($\kappa=-2$) avoids this conclusion while having no effect on the background evolution.
Still, it does not mean that the theory is necessarily free of all types of instabilities as
we will see shortly.

The other coefficients of Eq.~\eqref{eq:ST12} are given by
\begin{widetext}
 \begin{align}
  \mathcal{K}_T&=\frac{1}{2}(1+\chi_1)-\left(\frac{1}{2}+\frac{H^2}{\dot H}+\frac{H\ddot H}{\dot H^2}
  -\frac{H}{\dot H}\frac{\mathrm{d}}{\mathrm{d}t}\right)(\chi_1+\chi_2)~,\\
  \mathcal{M}_T&=\frac{1}{2}(1+\chi_1)-\left(\frac{1}{2}+\frac{H^2}{\dot H}-2\frac{\ddot H^2}{\dot H^3}+\frac{\dddot H}{\dot H^2}
  +2\frac{\ddot H}{\dot H^2}\frac{\mathrm{d}}{\mathrm{d}t}-\frac{1}{\dot H}\frac{\mathrm{d}^2}{\mathrm{d}t^2}\right)
  (\chi_1+\chi_2)~.
 \end{align}
 The criteria to avoid ghost and gradient instabilities are
 $\mathcal{K}_T>0$ and $\mathcal{M}_T>0$, respectively.
 By using the background EOMs (see Eq.~\eqref{eq:adot}, which can be rewritten as $H=\sqrt{V_1'/12}$,
 Eq.~\eqref{eq:chi1dot}, and Eq.~\eqref{eq:chi2dot} in the case for vacuum with $\varepsilon=0$),
 the conditions can be written solely in terms of the fields $\chi_1$, $\chi_2$ and their potentials
 $V_1(\chi_1)$, $V_2(\chi_2)$ as
 \begin{align}
  F_1:=&~\frac{V_2'{}^2}{V_1''}
   +\left[(\chi_1+\chi_2)-\frac{1}{4}(\chi_1+1)+\frac{1}{2}\frac{V_1+V_2}{V_1'}\right]\left(V_2'-(\chi_1+\chi_2)V_2''\right)
   +\frac{1}{4}(\chi_1+\chi_2)^2V_2''+\frac{1}{4}\frac{V_2'{}^2}{V_1'}(1-\chi_2)>0~,\label{con1}\\
   F_2:=&~
   -\frac{4 \left(V_1+V_2\right){}^2 \left(\chi _1+\chi _2\right) \left(V_2''\right){}^2}{V_2^3
   V_1'}-\frac{\left(\chi _1+\chi _2\right) \left(2 \chi _1+3 \chi _2-1\right){}^2 V_1'
   \left(V_2''\right){}^2}{V_2^3}\notag\\
   &+\frac{\left(2 \chi _1+3 \chi _2-1\right) \left(7 \chi _1+9
   \chi _2-2\right) V_1' V_2''}{2 V_2^2}+\frac{2 \left(6 \chi _1+7 \chi _2-1\right) V_1'
   V_2''}{V_2 V_1''}\notag\\
   &+\frac{2 V_2{}^{(3)} \left(V_1+V_2\right){}^2 \left(\chi _1+\chi_2\right)}{V_2^2 V_1'}-\frac{\left(5 \chi _1+8 \chi _2-3\right) V_1'}{2V_2}
   +\frac{V_2{}^{(3)} \left(\chi _1+\chi _2\right) \left(2 \chi _1+3 \chi _2-1\right){}^2V_1'}{2 V_2^2}\notag\\
   &-\frac{4 \left(V_1+V_2\right) \left[2 \chi _1^2+\left(5 \chi _2-1\right) \chi_1+\chi _2 \left(3 \chi _2-1\right)\right]
   \left(V_2''\right){}^2}{V_2^3}+\frac{\left(V_1+V_2\right) \left(11 \chi _1+15 \chi_2-4\right) V_2''}{V_2^2}\notag\\
   &+\frac{2 \left[2 \chi _1^2+\left(5 \chi _2-1\right) \chi _1+\chi _2
   \left(3 \chi _2-1\right)\right] V_2''}{V_2}+\frac{2 V_2{}^{(3)} \left(V_1+V_2\right) \left[2
   \chi _1^2+\left(5 \chi _2-1\right) \chi _1+\chi _2 \left(3 \chi_2-1\right)\right]}{V_2^2}\notag\\
   &+\frac{4 \left(V_1+V_2\right){}^2 V_2''}{V_2^2 V_1'}+\frac{8 V_2
   V_1{}^{(3)} V_1'}{\left(V_1''\right){}^3}-\frac{8 V_1'}{V_1''}+\frac{4 \left(V_1+V_2\right)
   V_2''}{V_2 V_1''}-\frac{4 V_2}{V_1''}-\frac{3 \left(V_1+V_2\right)}{V_2}+\frac{1}{2} \left(-8
   \chi _1-13 \chi _2+5\right)>0~.
\label{con2}
 \end{align}
\end{widetext}
These conditions will be tested for the different models of Secs.~\ref{sec:inflation} and \ref{sec:genesis}
shortly.

\subsubsection{Vector modes}

We shall consider vector fluctuations in the following gauge, where
\begin{equation}
\label{eq:defvector}
 g_{\mu\nu}^{(1)}\mathrm{d}x^\mu\mathrm{d}x^\nu=-\mathrm{d}t^2+2a\beta_i\mathrm{d}t\mathrm{d}x^i+a^2\delta_{ij}\mathrm{d}x^i\mathrm{d}x^j~.
\end{equation}
Here, the vector perturbations $\beta^i$ satisfy $\partial_i\beta^i=0$.
The Fourier components of $\beta_i$ are then defined by
\begin{equation}
 \beta_i=\int\frac{\mathrm{d}^3\vec{k}}{(2\pi)^3}\sum_{I=1,2}\left[\beta_{I}(\vec{k})e^I_i(\vec{k})\right]\mathrm{e}^{\mathrm{i}\vec{k}\cdot\vec{x}}~,
\end{equation}
where $\{e^1_i,e^2_i\}$ are orthogonal spatial vectors perpendicular to $\vec{k}$.
The second-order action for vector perturbations becomes
\begin{align}
 S_V^{(2)}=&~\frac{M_\mathrm{Pl}^2}{2}\int\mathrm{d}t\frac{\mathrm{d}^3\vec{k}}{(2\pi)^3}\sum_{I=1,2}a^3\frac{k^2}{a^2}
 \Big[\mathcal{G}_V\Big(\dot\beta_I^2-\frac{k^2}{a^2}\beta_I^2\Big)\nonumber\\
 &+\mathcal{K}_V\beta_I^2\Big]~,
\end{align}
where $\mathcal{G}_V=\mathcal{G}_T$ and $\mathcal{K}_V=\mathcal{K}_T$.
Accordingly, when $\kappa=-2$, which sets $\mathcal{G}_T=0$ to avoid Ostrogradski ghosts,
it turns out that $\mathcal{G}_V=0$ as well, and as a result, there are no dynamical vector modes.

\subsubsection{Scalar modes}

We shall then focus on the scalar fluctuations in the spatially flat gauge, where
$\chi_1=\chi_1^{(0)}+\delta\chi_1$, $\chi_2=\chi_2^{(0)}+\delta\chi_2$, and
\begin{equation}
\label{eq:defscalar}
 g_{\mu\nu}^{(1)}\mathrm{d}x^\mu\mathrm{d}x^\nu=-(1+2\Phi)\mathrm{d}t^2+2a\partial_iB\mathrm{d}t\mathrm{d}x^i+a^2\delta_{ij}\mathrm{d}x^i\mathrm{d}x^j~.
\end{equation}
The second-order action for scalar modes is then given by
\begin{widetext}
 \begin{equation}
  S_S^{(2)}=\frac{M_\mathrm{Pl}^2}{2}\int\mathrm{d}t\frac{\mathrm{d}^3\vec{k}}{(2\pi)^3}~a^3\left[
  \frac{4}{3}\frac{k^4}{a^4}\mathcal{G}_S(\Phi+a\dot B)^2
  +2\left(\frac{k^2}{a^2}aB-3H\Phi\right)\delta\dot\chi_2
  +M_{IJ}\Psi^I\Psi^J\right]~,
 \end{equation}
 where $\mathcal{G}_S=\mathcal{G}_T$ and\footnote{We
 omit the subscript $\vec{k}$ from the perturbation variables $\Psi^I$ when it is clear that they represent the Fourier components.}
 $\Psi^I:=(\Phi,B,\delta\chi_1,\delta\chi_2)$.
 The matrix $M_{IJ}$ is given by\footnote{As before, we omit the superscript $(0)$ for $\chi_1$ and $\chi_2$
 when it is understood that they represent background quantities.}
 \begin{equation}
  M_{IJ}=\left(
   \begin{array}{cccc}
    \frac{8H^2\left(\chi_1+\chi_2\right)}{\dot{H}}\frac{k^2}{a^2}+6H\left[H\left(2\chi_1
    +3\chi_2-1\right)+\dot{\chi}_2\right] & *  &*  & * \\
    -\frac{4H\left(\chi_1+\chi_2\right)}{3\dot{H}}\frac{k^4}{a^3}-\left(4\chi_1H
    +6\chi_2H-2H+\dot{\chi}_2\right)\frac{k^2}{a} & \frac{2k^4\left(2A_1+2\chi_1
    +3\chi_2-1\right)}{3a^2} & * & * \\
    -6H^2-\frac{V_1'}{2} & \frac{4Hk^2}{a} & -\frac{V_1''}{2} & 0 \\
    -9 H^2-\frac{V_2'}{2}-\frac{k^2}{a^2} & \frac{3 H k^2}{a} & 0 & -\frac{V_2''}{2} \\
   \end{array}
  \right)~,
 \end{equation}
\end{widetext}
where $*$ stands for symmetric components.
Since no time derivatives of $\Phi$, $B$, and $\delta \chi_1$ appear in the second-order action with $\kappa=-2$,
these modes are nondynamical. Then, these variables can be eliminated by their equations of motion.
After removing the nondynamical modes, the resulting action can be written solely in term of $\delta\chi_2$ as follows:
\begin{equation}
\label{eq:SS2I}
 S^{(2)}_S=\frac{M_{\mathrm{Pl}}^2}{2}\int\mathrm{d}t\frac{\mathrm{d}^3\vec{k}}{(2\pi)^3}~
 a^3\left[{\cal K}_{S}(\delta\dot{\chi}_2)^2-{\cal M}_{S}(\delta\chi_2)^2\right]~,
\end{equation}
where
\begin{align}
 {\cal K}_S=&~12{\cal K}_T\Big[\Big(\frac{4}{3}\frac{k^2}{a^2}\frac{(\chi_1+\chi_2)}{\dot{H}}+\frac{\dot{\chi}_2}{H}\Big)^2 \nonumber \\
  &-8\mathcal{K}_T\Big(\frac{4}{3}\frac{k^2}{a^2}\frac{(\chi_1+\chi_2)}{\dot{H}}+\frac{2H\dot{\chi}_1}{\dot{H}}
  +\frac{\dot{\chi}_2}{H} \nonumber \\
  &+2\chi_1+3\chi_2-1\Big)\Big]^{-1}~, \\
 {\cal M}_S=&~\frac{{\cal K}_S^2}{{\cal K}_T^2}\left(C_8\frac{k^8}{a^8}+C_6\frac{k^6}{a^6}+C_4\frac{k^4}{a^4}+C_2\frac{k^2}{a^2}+C_0\right)~,
 \label{eq:calMS}
\end{align}
with
\begin{align}
 C_8=&~192H^2\dot{H}^2(\chi_1+\chi_2)^2\frac{H}{\dot{\chi}_2}\Big[4H\dot{H}\dot{\chi}_2(\chi_1+\chi_2) \nonumber \\
 &-4H\ddot{H}(\chi_1+\chi_2)^2-\dot{H}\dot{\chi}_1\dot{\chi}_2\Big]~.
\end{align}
We do not write down the form of the other $C_n$ coefficients because
they are not so relevant in the following stability analysis.

In the small scale limit ($k/a \rightarrow\infty$), one finds
\begin{align}
 {\cal K}_S\simeq&~A_S{\cal K}_T, \\
 {\cal M}_S\simeq&~B_S{\cal K}_S\frac{H}{\dot{\chi}_2}\Big[4H\dot{H}\dot{\chi}_2(\chi_1+\chi_2)-4H\ddot{H}(\chi_1+\chi_2)^2 \nonumber \\
 &-\dot{H}\dot{\chi}_1\dot{\chi}_2\Big]~,
\end{align}
where $A_S$ and $B_S$ are simply two real positive constants.
Thus, the ghost instability in the scalar sector is avoidable when it is absent in the tensor sector,
i.e.~when ${\cal K}_T>0$ is satisfied.
In addition, the gradient instability is absent when ${\cal M}_S>0$, so when
\begin{align}
 F_3:=&~\frac{H}{\dot{\chi}_2}\Big[4H\dot{H}\dot{\chi}_2(\chi_1+\chi_2)-4H\ddot{H}(\chi_1+\chi_2)^2 \nonumber \\
 &-\dot{H}\dot{\chi}_1\dot{\chi}_2\Big]>0~.\label{grascalar}
\end{align}
By using the vacuum background EOMs, this condition can be rewritten as
\begin{align}
 F_3=(\chi_1+\chi_2)[(\chi_1+\chi_2)V_2''-V_2']-\frac{(V_2')^2}{V_1''}>0~.\label{con3}
\end{align}

\begin{figure*}
 \centering
 \includegraphics[width=0.45\textwidth]{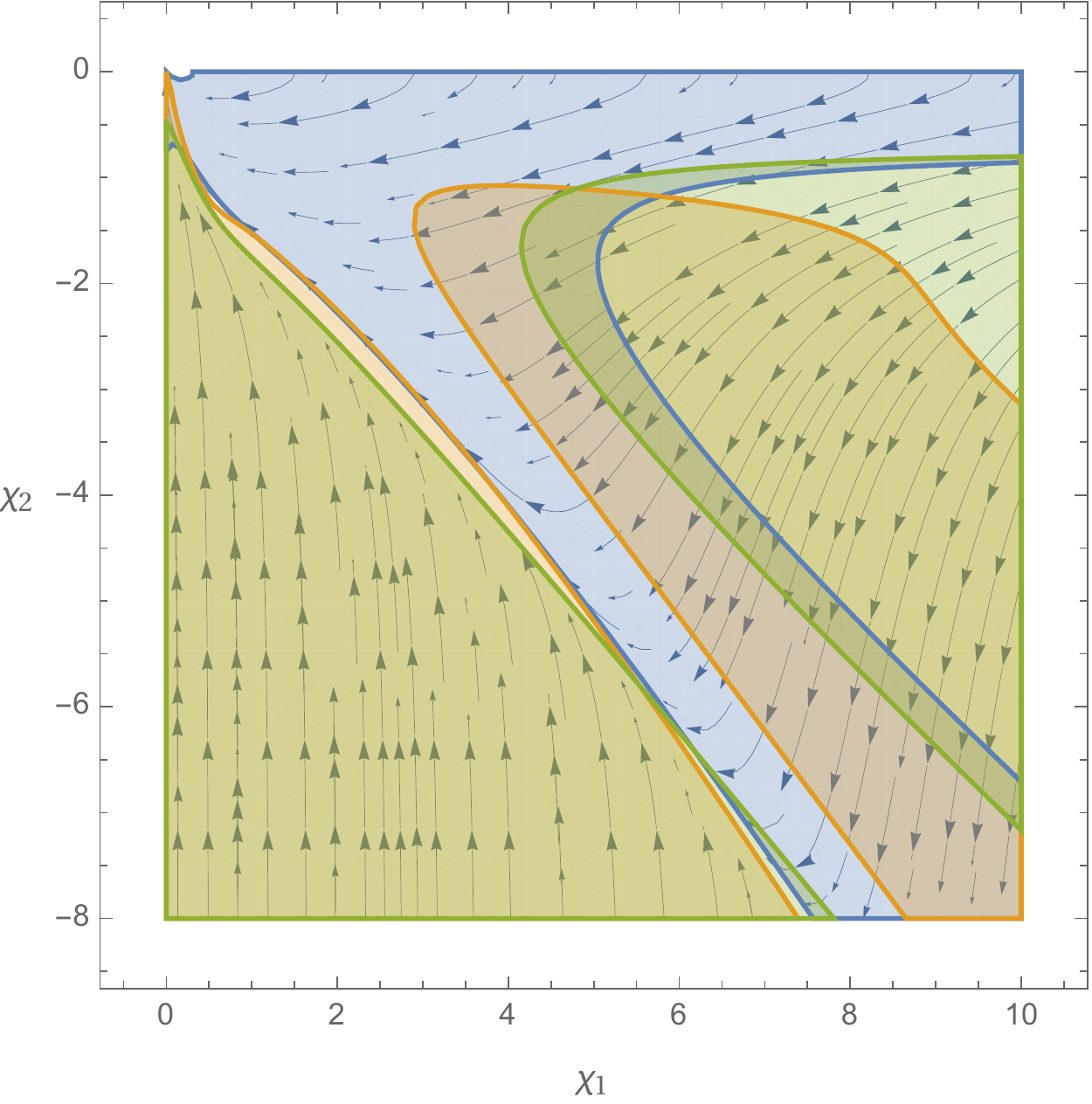}\hfill
 \includegraphics[width=0.45\textwidth]{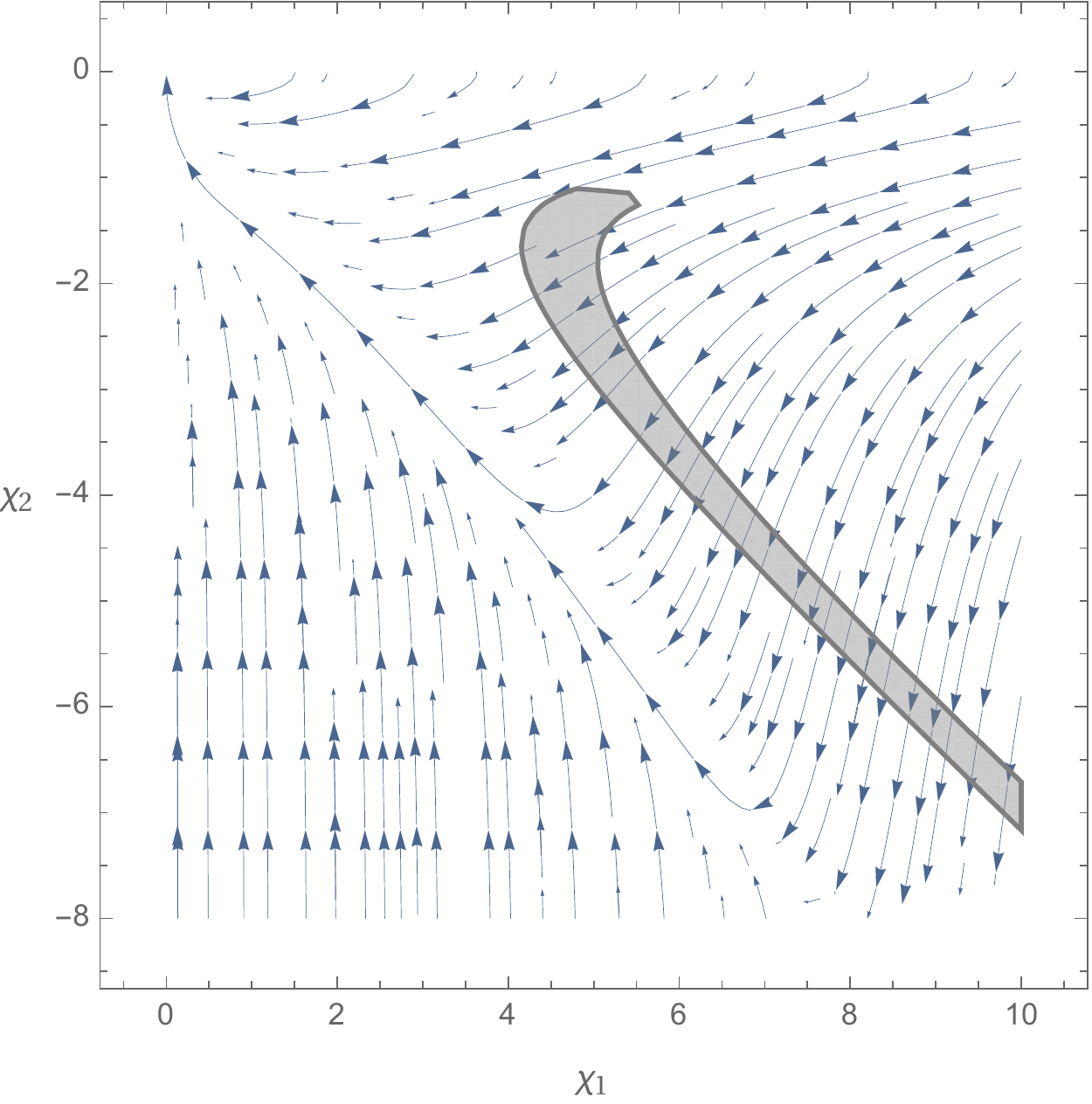}
 \caption{Phase-space diagram of the inflationary model of Sec.~\ref{sec:inflation}. On top of the phase-space trajectories,
 we show the regions that are stable according to the conditions that were derived in the text.
 In particular, in the left-hand plot, the blue, orange, and green regions show where the conditions
 $F_1>0$ [Eq.~\eqref{con1}], $F_2>0$ [Eq.~\eqref{con2}], and $F_3>0$ [Eq.~\eqref{con3}], respectively, are satisfied.
 The gray shaded area in the right-hand plot shows where all three conditions are met at the same time.}
\label{fig:inflationstability}
\end{figure*}

Regarding gradient instabilities which can occur on sub-Hubble scales ($k/a\gg H$),
the general procedure is to check for the stability of modes within the validity range of the EFT,
i.e.~for $H\ll k/a\leq\Lambda_\mathrm{cut}$ (see for instance Ref.\ \cite{Koehn:2015vvy}),
where for our models $H\leq H_\mathrm{max}\leq\Lambda_\mathrm{cut}$.
The condition given by Eq.\ \eqref{con3} can be viewed as the one which ensures
that the shortest wavelength modes ($k/a\sim\Lambda_\mathrm{cut}$) do not suffer from gradient instabilities.
However, since the perturbed action exhibits a modified dispersion relation,
i.e., since Eq.\ \eqref{eq:calMS} has terms of order $k^2$, $k^4$, $k^6$, and $k^8$,
longer wavelength modes (longer than $\Lambda_\mathrm{cut}^{-1}$, but still smaller than $H^{-1}$)
could still suffer from gradient instabilities.
As long as the duration for such gradient instabilities is not too long though, their amplification remains controllable
in comparison to the smaller wavelength modes which easily blow up
(within a time scale of the order of $\Lambda_{\text{cut}}^{-1}$).
This is why we only focus on the stability of the shortest wavelength modes.

\subsubsection{Stability analysis}

In summary, with $\kappa=-2$, we saw that there is no Ostrogradski instability.
Then, we derived three conditions given by Eq.~\eqref{con1}, which determines when the model is free of ghost instabilities
in the tensor and scalar sectors, and Eqs.\ \eqref{con2} and\ \eqref{con3}, which determine
when the model is free of gradient instabilities in the tensor and scalar sector, respectively.
The conditions depend on the potentials $V_1(\chi_1)$ and $V_2(\chi_2)$ and on the field values $\chi_1$ and $\chi_2$,
so we need to study specific models to comment on the stability of the given theory.

\begin{figure*}
 \centering
 \includegraphics[width=0.45\textwidth]{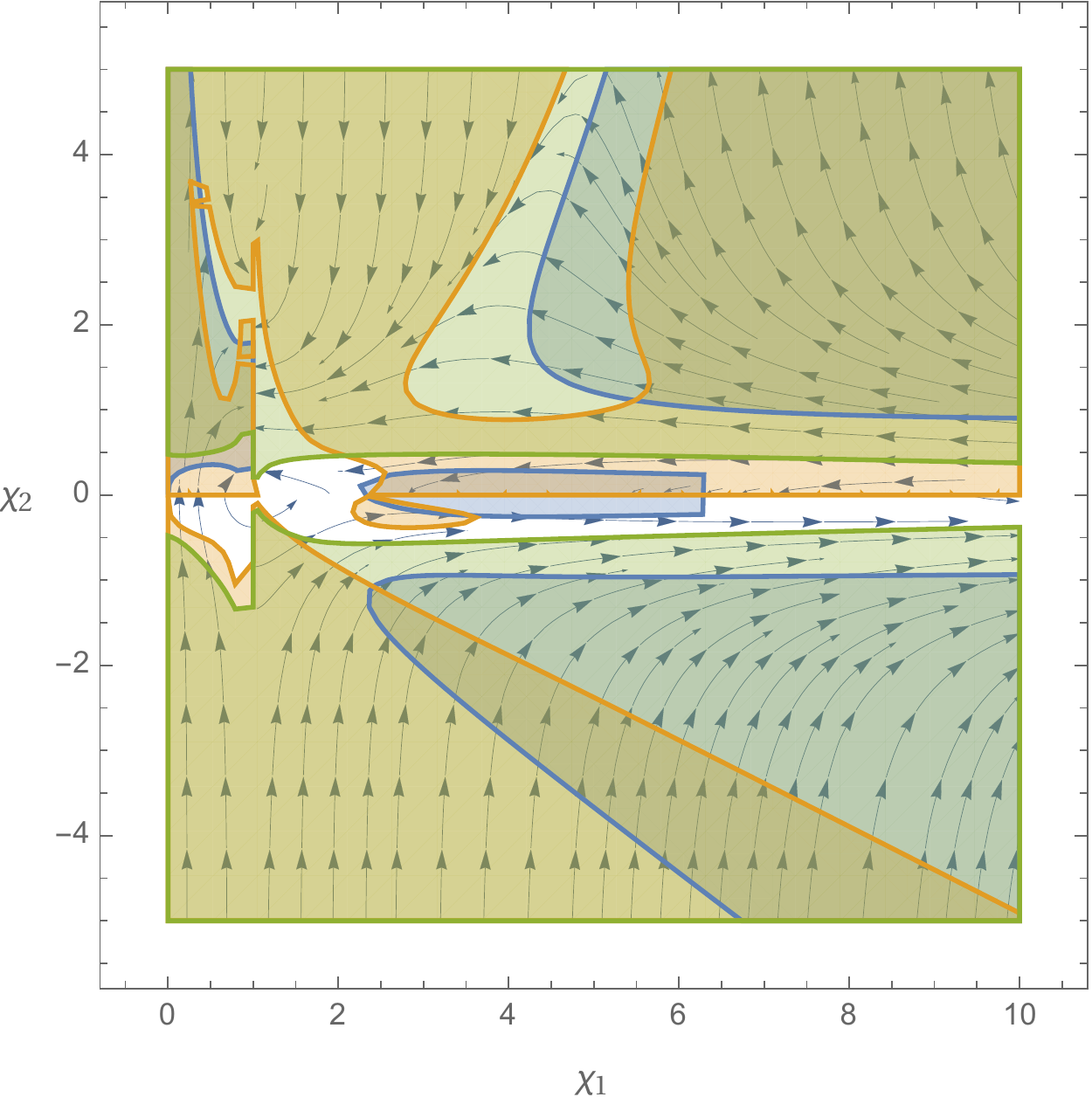}\hfill
 \includegraphics[width=0.45\textwidth]{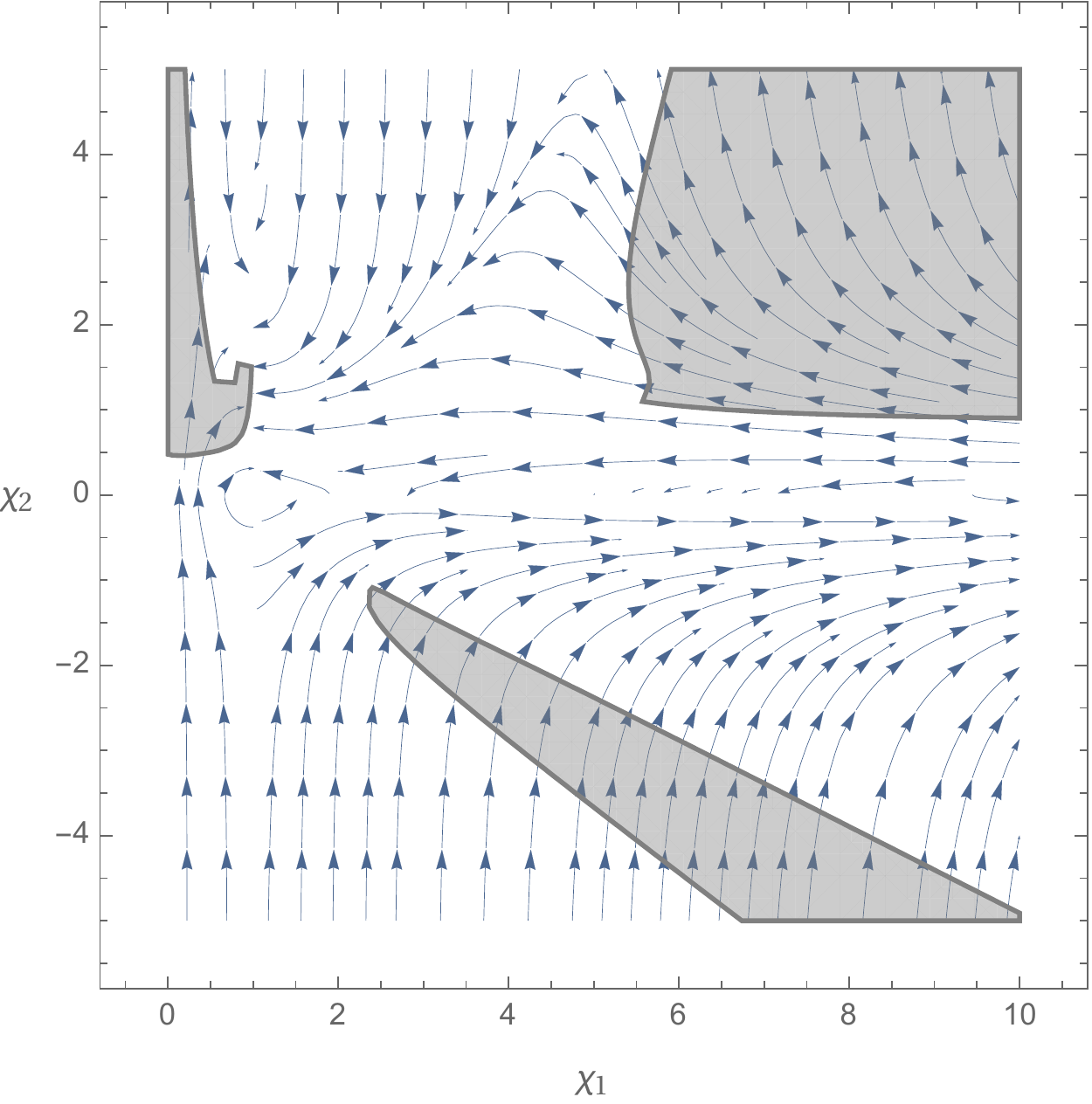}
 \caption{Phase-space diagram of the genesis model of Sec.\ \ref{sec:genesis}.
 The convention used to illustrate the stable regions is described in Fig.~\ref{fig:inflationstability}.}
\label{fig:genesisstability}
\end{figure*}

Starting with the inflationary model of Sec.\ \ref{sec:inflation}, where the potentials are given by Eqs.\ \eqref{eq:V1inf}
and\ \eqref{eq:V2inf}, we plot the regions of phase space that satisfy the three conditions in Fig.~\ref{fig:inflationstability}.
The individual conditions are shown in the left-hand plot, and we see that there are large regions of phase space that
can avoid ghost or gradient instabilities in the tensor or scalar sectors.
However, when we look at the right-hand plot, which shows the combined region where all stability conditions are met,
we see that there is, actually, only a very small region of phase space that is not unstable.
In particular, the trajectories that correspond to asymptotically de Sitter and Minkowski (e.g., the green curve
in Fig.~\ref{fig:inflationtrajectories}) are generally unstable throughout their evolution, except in a very small region of phase space.

We show the same types of plots in Fig.~\ref{fig:genesisstability} for the genesis scenario of Sec.~\ref{sec:genesis}.
There as well, there are large regions of phase space that can avoid ghost and gradient instabilities in the tensor or scalar sectors,
but it remains that only small portions of those can be stable with regards to all types of instabilities at the same time.
In particular, the interesting trajectories (e.g., the black and purple curves of Fig.~\ref{fig:genesistrajectories})
are unstable throughout their evolution.

\subsection{Equivalence with $f(R,\mathcal{G})$ gravity}\label{sec:equivGaussBonnet}

We have shown that the Weyl square term with $\kappa=-2$ kills the dangerous degrees of freedom and can relax the instability around
a flat FLRW background spacetime. This result can be understood because this theory is included by $f(R,\mathcal{G})$ gravity,
whose cosmological solutions are perturbatively stable. However, it is shown by Ref.\ \cite{DeFelice:2010hg} that the perturbations of flat
FLRW spacetime in $f(R,\mathcal{G})$ gravity lose a degree of freedom, which reappears as a ghost around anisotropic spacetimes.
Therefore, our theory necessarily suffers from the same instability.

Let us demonstrate the equivalence between the above model and $f(R, {\cal G})$ gravity. 
The Gauss-Bonnet term ${\cal G}$ is defined by
\begin{align}
{\cal G}:= R^2-4R_{\mu\nu}R^{\mu\nu}+R_{\mu\nu\rho\sigma}R^{\mu\nu\rho\sigma}~.
\end{align}
Plugging the relation 
\begin{align}
 R_{\mu\nu\rho\sigma}R^{\mu\nu\rho\sigma}=C_{\mu\nu\rho\sigma}C^{\mu\nu\rho\sigma}+2R_{\mu\nu}R^{\mu\nu}-\frac{1}{3}R^2
\end{align}
into the  definition of the Gauss-Bonnet term, we obtain
\begin{align}
 {\cal G}=\frac{2}{3}R^2-2R_{\mu\nu}R^{\mu\nu}+C_{\mu\nu\rho\sigma}C^{\mu\nu\rho\sigma}~.
\end{align}
Then, we note that the second curvature-invariant function of Eq.~\eqref{eq:Imodels12} can be written as (with $\kappa=-2$)
\begin{align}
 I_2=\sqrt{12R_{\mu\nu}R^{\mu\nu}-3R^2-6C_{\mu\nu\rho\sigma}C^{\mu\nu\rho\sigma}}=\sqrt{R^2-6{\cal G}}~.
\end{align}
Accordingly, we see that $I_2$ and $I_1$ (which are given by $I_1=I_2+R=\sqrt{R^2-6{\cal G}}+R$)
are functions of $R$ and ${\cal G}$. 
Thus, the solutions of the EOMs for $\chi_1$ and $\chi_2$ can be written as 
\begin{equation}
 \chi_1=\chi_1(R,{\cal G})~,\qquad\chi_2=\chi_2(R,{\cal G})~,
\end{equation}
and by plugging these solutions into Eq.~\eqref{eq:masteraction},
the original action becomes only a function of $R$ and the Gauss-Bonnet term ${\cal G}$, i.e.~
\begin{align}
 S=\int\mathrm{d}^4x~\sqrt{-g}f(R,{\cal G})~.
\end{align}
In conclusion, this theory is a specific model of $f(R, {\cal G})$ gravity.

\section{New nonsingular model with a new curvature scalar}\label{sec:model3}

\subsection{Setup}\label{sec:setupmodel3}

Let us investigate other curvature scalars, which reduce to $H$ and $\dot{H}$ at the background level.
In particular, we focus our attention on curvature scalars that are functions of the Ricci scalar and its derivatives.

Let us consider the following tensor constructed from the first derivative of $R$,
\begin{align}
  X^{\mu}{}_{\nu}:=g^{\mu\rho}\nabla_{\rho}R\nabla_{\nu}R~.
\end{align}
For a flat FLRW background, this quantity reduces to
\begin{equation}
 X^{\mu}{}_{\nu}^{(0)}=-\left[6(4H\dot{H}+\ddot{H})\right]^2\mathrm{diag}(1,0,0,0)~.
\end{equation}
The trace of the tensor $X^\mu{}_\nu$ is
\begin{equation}
 X=X^\mu{}_\mu=\nabla_\mu R\nabla^\mu R~,
\end{equation}
and at the background level, it reduces to
\begin{equation}
 X^{(0)}=-\left[6(4H\dot{H}+\ddot{H})\right]^2~.
\end{equation}
Since the second time derivative of the Hubble parameter appears here, we need to consider another curvature scalar to remove $\ddot{H}$.
Thus, let us consider the second derivative of $R$,
\begin{equation}
 {\cal R}^{\mu}{}_{\nu}:=\nabla^{\mu}\nabla_{\nu}R~,
\end{equation}
whose expression in a FLRW spacetime reduces to
\begin{equation}
 {\cal R}^{\mu}{}_{\nu}^{(0)}=-24H\mathrm{diag}\left[\frac{\dot{H}^2}{H}+\ddot{H}+\frac{\dddot{H}}{4H},
 \Big(H\dot{H}+\frac{\ddot{H}}{4}\Big)\delta^i{}_j\right]~.
\end{equation}
Since $\dddot{H}$ appears only in the $(0,0)$ component,
it can be removed by the tensor defined by
\begin{equation}
 P^\mu{}_\nu:=\delta^\mu{}_\nu-\frac{X^\mu{}_\nu}{X}~,
\end{equation}
whose FLRW limit now is $P^\mu{}_\nu^{(0)}=\mathrm{diag}(0,1,1,1)$.
In fact, we can construct the scalar quantity
$\mathrm{tr}[P\mathcal{R}P\mathcal{R}]$, which gives
\begin{equation}
 \mathrm{tr}[P\mathcal{R}P\mathcal{R}]^{(0)}=-3XH^2~,
\end{equation}
at the background level.
Therefore, the following quantity is a good candidate as a limiting curvature-invariant function,
\begin{align}
\label{eq:I12}
 I_1:=&~-12\frac{\mathrm{tr}[P{\cal R}P{\cal R}]}{3X} \\
 =&~-\frac{4}{X^3}\Big[X^2(\nabla_{\mu}\nabla_{\nu}R)(\nabla^{\mu}\nabla^{\nu}R) \nonumber \\
 &-2X(\nabla^{\mu}R\nabla^{\nu}R)(\nabla_{\mu}\nabla_{\rho}R)(\nabla_{\nu}\nabla^\rho R) \nonumber \\
 &+(\nabla^{\mu}R \nabla^{\nu}R \nabla_{\mu}\nabla_{\nu}R)^2\Big]\\
 =&~-\frac{1}{X^3}\left[4X^2(\nabla\nabla R)^2-2X(\nabla X)^2+(\nabla R\nabla X)^2\right]~,
\end{align}
which satisfies
\begin{equation}
 I_1^{(0)}=12H^2~,
\end{equation}
as required.
In order to bound $\dot{H}$, we introduce as before
\begin{equation}
\label{eq:I22}
 I_2:=I_1-R~;
\end{equation}
hence
\begin{equation}
  I_2^{(0)}=-6\dot{H}~,
\end{equation}
again as required.

\subsection{Cosmological perturbations and stability analysis}\label{sec:stabilitymodel3}

\subsubsection{Tensor modes}

By substituting the definition of tensor perturbations [Eq.~\eqref{eq:deftensor}],
the second-order action for tensor modes can be obtained as
\begin{align}
 S_T^{(2)}=&~\frac{M_{\mathrm{Pl}}^2}{2}\int\mathrm{d}t\frac{\mathrm{d}^3\vec{k}}{(2\pi)^3}\sum_{I={+,\times}}a^3
 \Big({\cal K}_{T}\dot{h}^{I}_{\vec{k}}\dot{h}^{I}_{-\vec{k}} \nonumber \\
 &-{\cal M}_{T}\frac{k^2}{a^2}h^I_{\vec{k}}h^{I}_{-\vec{k}}\Big)~,
\end{align}
where
\begin{align}
 &{\cal K}_T=\frac{1+4\chi_1+3\chi_2}{2}~,\\
 &{\cal M}_T=\frac{1-\chi_2}{2}~.
\end{align}
Thus, the tensor sector does not include higher-derivative terms, so the number of degrees of freedom is 2,
and no Ostrogradski instabilities appear.
The stability conditions for ghost and gradient instabilities in the tensor sector are then given by
\begin{align}
\label{eq:F1_2}
 &F_1:=1+4\chi_1+3\chi_2>0~, \\
\label{eq:F2_2}
 &F_2:=1-\chi_2>0~.
\end{align}

\subsubsection{Vector modes}

Using Eq.~\eqref{eq:defvector}, the second-order action for vector modes can be derived as
\begin{equation}
 S^{(2)}_V=\frac{M_{\mathrm{Pl}}^2}{2}\int\mathrm{d}t\frac{\mathrm{d}^3\vec{k}}{(2\pi)^3}a^3\sum_{I=1,2}\frac{k^2}{a^2}{\cal K}_{V}\beta_{I}^2,
\end{equation}
with $\mathcal{K}_V=\mathcal{K}_T$.
Therefore, no vector modes exist in this theory.

\subsubsection{Scalar modes}

Let us investigate the scalar perturbations defined in Eq.~\eqref{eq:defscalar}.
Let us introduce a new perturbation variable $\varphi$ in terms of $\Phi$ and $B$ by the equation
\begin{align}
\label{varphi=}
 \varphi_{\vec{k}}=\Phi_{\vec{k}}-\frac{k^2}{3aH}B_{\vec{k}}~,
\end{align}
and we regard the components of $\Psi^I:=(B,\delta\chi_1,\delta\chi_2,\varphi)$ as independent variables.

\begin{figure*}
 \centering
 \includegraphics[width=0.45\textwidth]{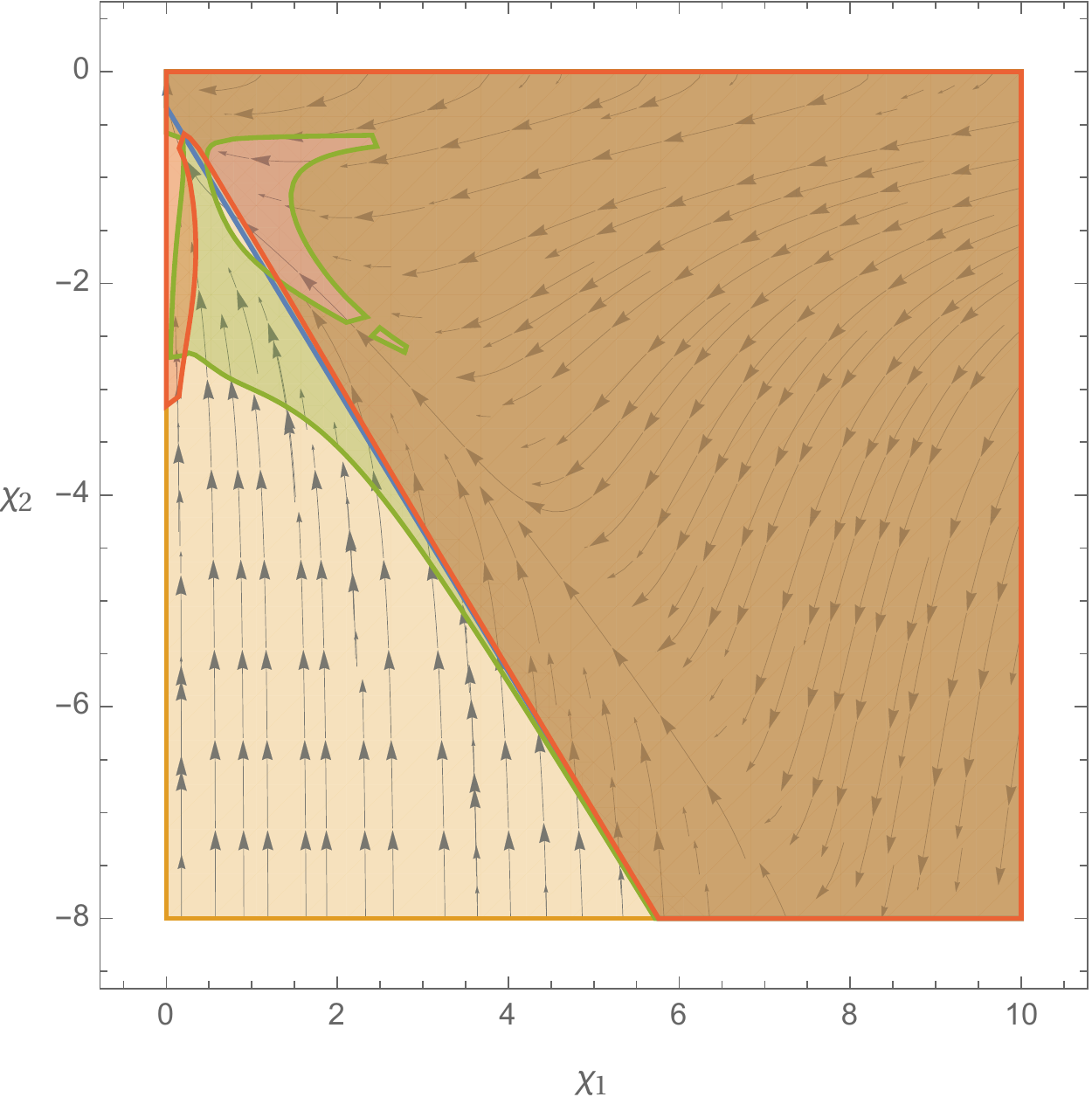}\hfill
 \includegraphics[width=0.45\textwidth]{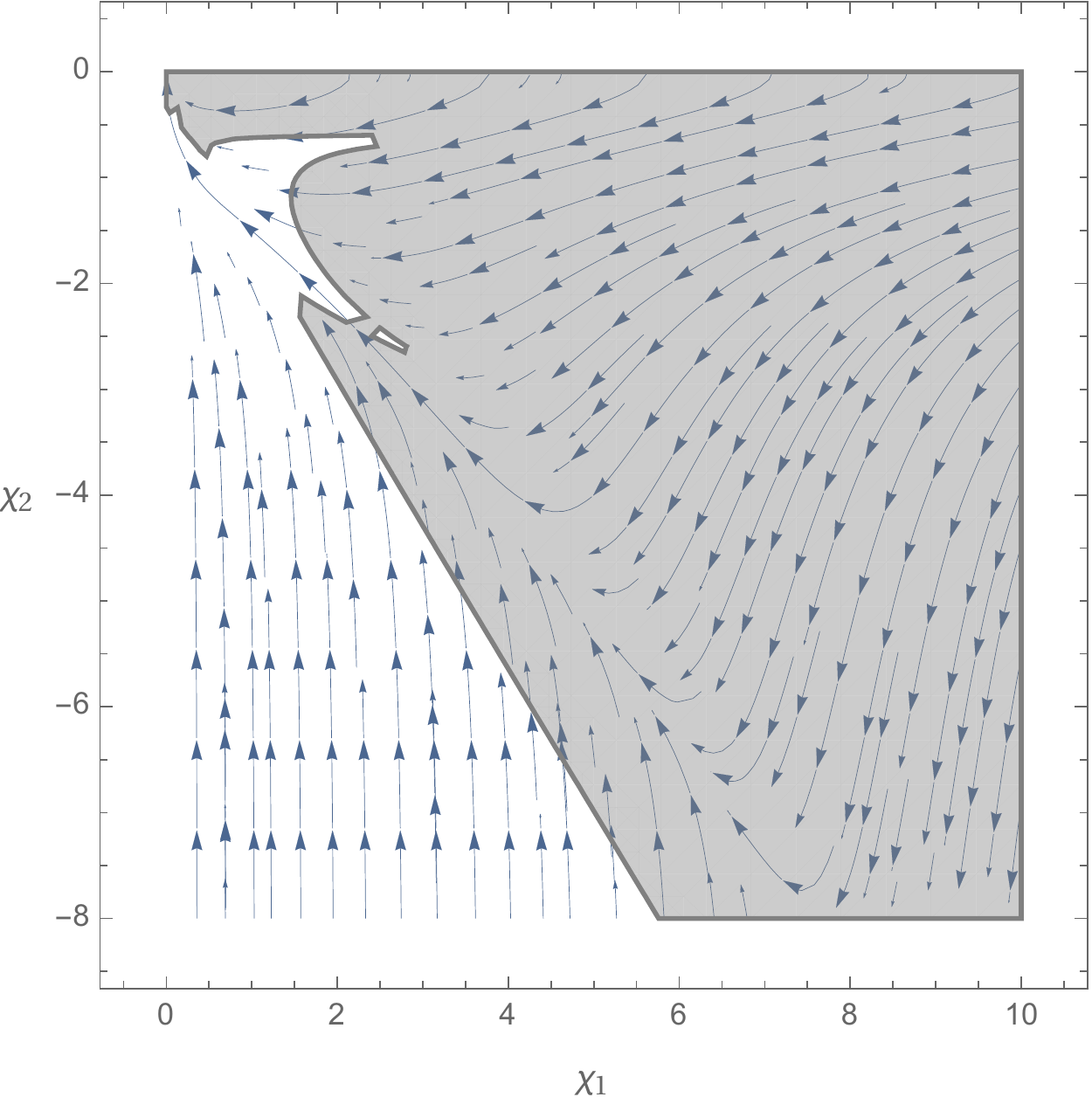}
 \caption{Phase-space diagram of the inflationary model of Sec.~\ref{sec:inflation}. On top of the phase-space trajectories,
 we show the regions that are stable according to the conditions that were derived in the text.
 In particular, in the left-hand plot, the blue, orange, green, and red regions show where the conditions
 $F_1>0$ [Eq.~\eqref{eq:F1_2}], $F_2>0$ [Eq.~\eqref{eq:F2_2}], $F_3>0$ [Eq.~\eqref{eq:F3_2}], and $F_4>0$ [Eq.~\eqref{eq:F4_2}],
 respectively, are satisfied. The gray shaded area in the right-hand plot shows where all four conditions are met at the same time.}
\label{fig:inflationstability2}
\end{figure*}

The second-order action for scalar modes can then be calculated as
\begin{equation}
 S^{(2)}_S=\frac{M_\mathrm{Pl}^2}{2}\int\mathrm{d}t\frac{\mathrm{d}^3\vec{k}}{(2\pi)^3}
 a^3\left[K\dot{\varphi}^2+L_I\Psi^I\dot{\varphi}+M_{IJ}\Psi^I\Psi^J\right]~,
\end{equation}
where we omit writing down the specific form of the terms $K$, $L_I$, and $M_{IJ}$.
Since the action does not include the time derivative of 
$B$, $\delta \chi_1$, and $\delta \chi_2$,
we can remove these variables by the EOMs.
The resulting action is given by
\begin{equation}
 S_S^{(2)}=\frac{M_\mathrm{Pl}^2}{2}\int\mathrm{d}t\frac{\mathrm{d}^3\vec{k}}{(2\pi)^3}~a^3\Big({\cal K}_S\dot{\varphi}^2-{\cal M}_S\varphi^2\Big)~,
\end{equation}
where the coefficients, in the limit where $k/a \rightarrow\infty$, are given by
\begin{align}
 {\cal K}_S=&~\frac{a^4}{k^4}F_3+{\cal O}\Big[\Big(\frac{k}{a}\Big)^{-6}\Big]~,\\
 {\cal M}_S=&~{\cal K}_S\frac{k^2}{a^2}F_4+{\cal O}\Big[\Big(\frac{k}{a}\Big)^0\Big]~, 
\end{align}
with
\begin{widetext}
 \begin{align}
 \label{eq:F3_2}
  F_3=&-\frac{3V_1'}{8[3(\chi_1+\chi_2)V_1''V_2''-2V_1'(V_1''-V_2'')]}
   \Big[2(4\chi_1+3\chi_2+1)(V_1')^2(V_1''-V_2'')+12(\chi_1+\chi_2)(V_2')^2V_1'' \nonumber \\
   &+V_1'\left\{-16(\chi_1+\chi_2)V_2'V_1''-\left[4\chi_1^2+(5\chi_2+3)\chi_1+\chi_2(\chi_2+3)\right]V_1''V_2''+8(V_2')^2\right\}\Big]~, \\
 \label{eq:F4_2}
  F_4=&-\frac{(\chi_2-1)\left[3(\chi_1+\chi_2)V_1''V_2''+2V_1'(V_1''+V_2'')\right]}{2(4\chi_1+3\chi_2+1)V_1'(V_1''+V_2'')
   +\left[4\chi_1^2+(5\chi_2+3)\chi_1+\chi_2(\chi_2+3)\right]V_1''V_2''}~.
 \end{align}
\end{widetext}

\subsubsection{Stability analysis}

\begin{figure*}
 \centering
 \includegraphics[width=0.45\textwidth]{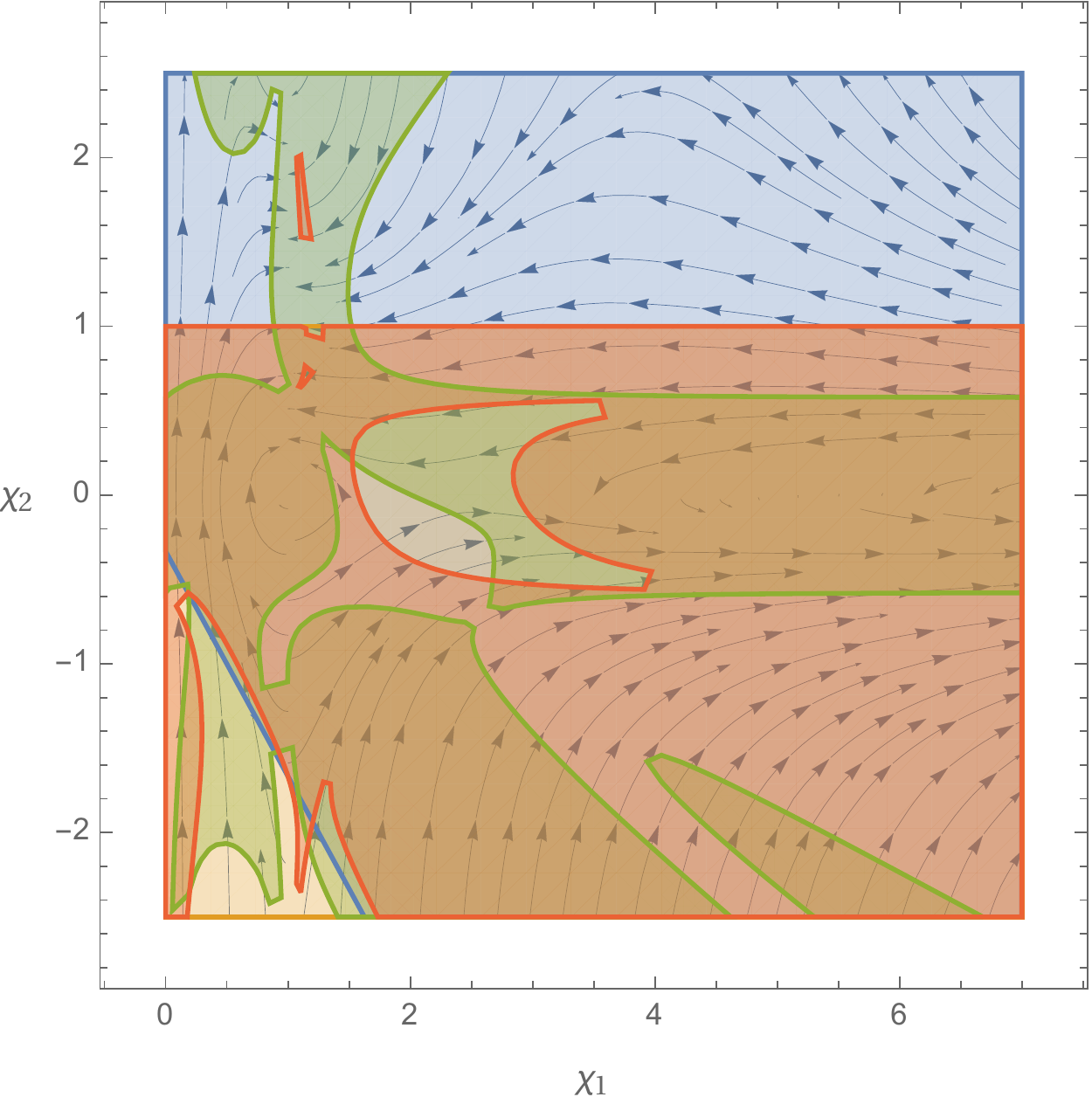}\hfill
 \includegraphics[width=0.45\textwidth]{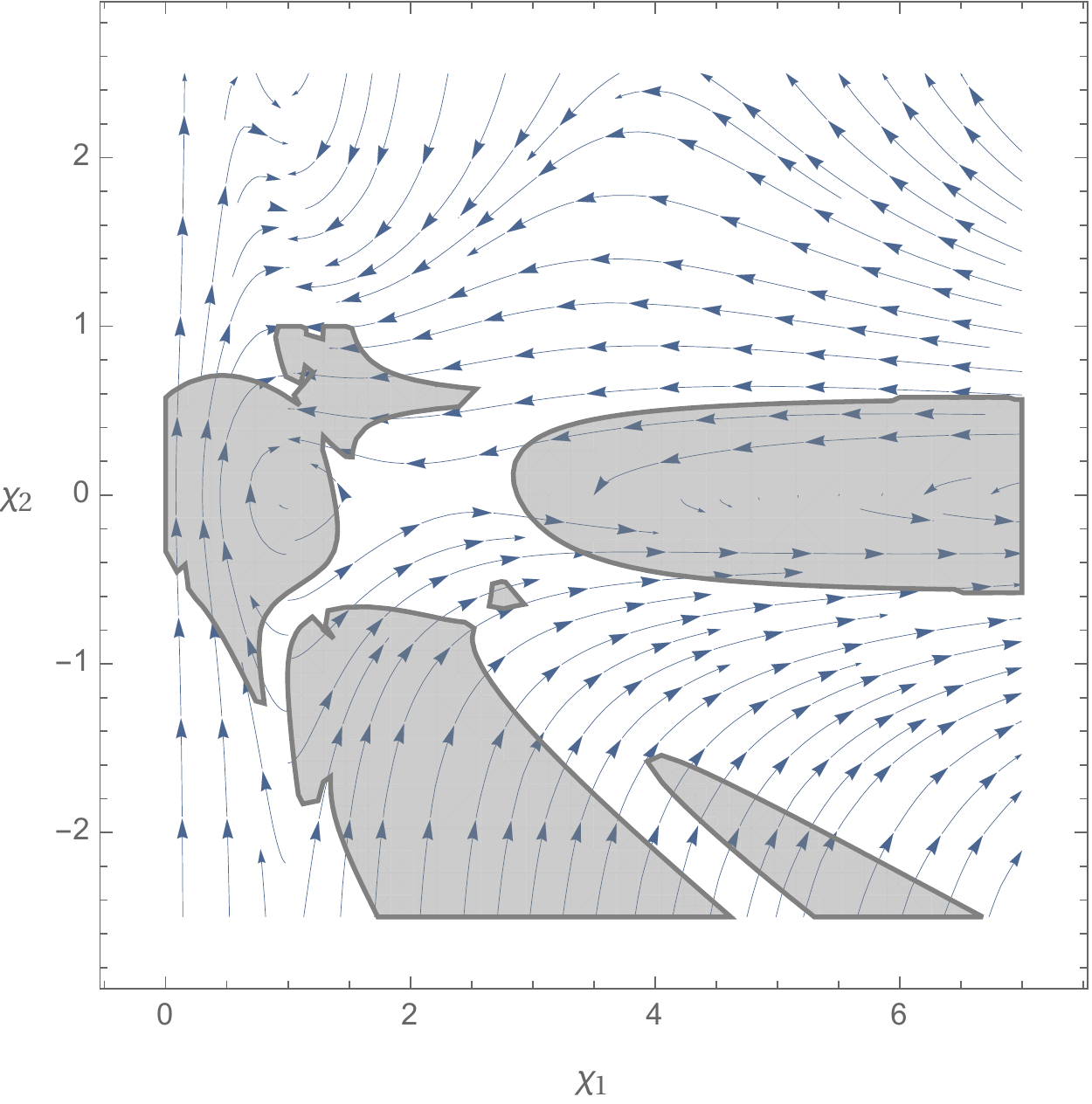}
 \caption{Phase-space diagram of the genesis model of Sec.\ \ref{sec:genesis}.
 The convention used to illustrate the stable regions is described in Fig.~\ref{fig:inflationstability2}.}
\label{fig:genesisstability2}
\end{figure*}

In summary, we derived four conditions given by Eqs.~\eqref{eq:F1_2}-\eqref{eq:F2_2}
and Eqs.~\eqref{eq:F3_2}-\eqref{eq:F4_2},
which determine when the model is free of ghost and gradient instabilities
in the tensor and scalar sectors, respectively.
Once again, the conditions depend on the potentials $V_1(\chi_1)$ and $V_2(\chi_2)$ and on the field values $\chi_1$ and $\chi_2$,
so we need to study specific models to comment on the stability of the given theory.

Starting with the inflationary model of Sec.\ \ref{sec:inflation},
we plot the regions of phase space that satisfy the four conditions in Fig.~\ref{fig:inflationstability2}.
Once again, there are large regions of phase space that can avoid ghost or gradient instabilities in the tensor or scalar sectors.
The major difference with Fig.~\ref{fig:inflationstability} is apparent in the right-hand plot of Fig.~\ref{fig:inflationstability2},
which shows the combined region where all stability conditions are met.
Indeed, whereas the previous theory had only a small strip of phase space that was stable,
there is now a large portion of the phase space that is free of all types of instabilities.
Furthermore, it is precisely in this region that we obtained the interesting background trajectories
starting from de Sitter spacetime and ending in Minkowski spacetime (e.g., the green curve in Fig.~\ref{fig:inflationtrajectories}).
For this class of trajectories, there only seems to be a small region of phase space
around $0\lesssim\chi_1\lesssim 2$ and $-0.5\lesssim\chi_2\lesssim-2.5$
where there still appears to be some instability.
Yet, it seems clear that the curvature invariants given by Eqs.~\eqref{eq:I12} and\ \eqref{eq:I22}
lead to a much more stable theory compared to the curvature invariants of Eq.~\eqref{eq:Imodels12}
that were analyzed in Sec.~\ref{sec:stabilitymodels1and2}.

For the genesis model of Sec.~\ref{sec:genesis}, the stable phase-space regions are shown in Fig.~\ref{fig:genesisstability2}.
There as well, there now are larger regions of phase space that can avoid all types of instabilities.
In particular, the interesting trajectories (e.g., the black and purple curves of Fig.~\ref{fig:genesistrajectories})
are \emph{stable} throughout their evolution. This reinforces our conclusion that the limiting curvature theory with
curvature invariants \eqref{eq:I12} and\ \eqref{eq:I22} is more stable
and leads to interesting nonsingular cosmological scenarios that can remain stable throughout their evolution.

\subsection{Stability around an anisotropic background}

At this point, the new curvature-invariant function leads to generally
stable nonsingular solutions around isotropic backgrounds. Still, one
may worry that this new theory might still possess undesired
ghosts. Indeed, it appears that the new curvature scalar of this section
is not included in the ghost free theories found in Ref.\ \cite{Naruko:2015zze},
which can be further mapped into multifield extensions of Horndeski theories\ \cite{Padilla:2012dx,Kobayashi:2013ina,Ohashi:2015fma}.

One situation where new ghosts may appear is around anisotropic backgrounds.
In the model of Sec.~\ref{sec:models1and2}, which we showed to be equivalent to a $f(R, {\cal G})$ theory of gravity,
a degree of freedom is lost around a FLRW background and such a mode reappears as a ghost once the background is deformed to an
anisotropic Bianchi type I universe. The purpose of this subsection is to investigate the stability of our new theory
against such a nonperturbative deformation of the background spacetime.
We shall investigate the perturbations around a Bianchi type I universe with the rotational symmetry in the $y-z$ plane,
so the background metric is given by
\begin{align}
 g^{(0)}_{\mu\nu}\mathrm{d}x^\mu\mathrm{d}x^\nu =& - \mathrm{d}t^2 + a(t)^2 \Big[\mathrm{e}^{-4 \delta(t)} \mathrm{d}x^2 \nonumber \\
 &+ \mathrm{e}^{2 \delta(t)}(\mathrm{d}y^2 + \mathrm{d}z^2) \Big]~.\label{BianchiI}
\end{align}
At the background level, the curvature scalar invariants now include shear terms,
\begin{align}
 &I_1^{(0)} = 12H^2 + 24 \sigma^2~, \\
 &I_2^{(0)} = -6 \dot{H} + 18\sigma^2~,
\end{align}
where the shear $\sigma$ is defined by
\begin{align}
 \sigma(t) := \dot{\delta}(t)~,
\end{align}
and so, the background dynamics is different from that analyzed in Sec.\ \ref{sec:setupmodel1and2}.

We focus on whether or not additional degrees of freedom of perturbations appear independently of the background dynamics.
One would usually start with the action given by Eq.\ \eqref{eq:masteraction} with curvature invariants given by Eqs.\ \eqref{eq:I12}
and\ \eqref{eq:I22} and perturb the action according to the scalar-vector-tensor decomposition of the perturbed
metric given by Eqs.\ \eqref{eq:deftensor},\ \eqref{eq:defvector}, and\ \eqref{eq:defscalar}.
Following the methodology of\ \cite{Watanabe:2010fh} for perturbations around an anisotropic background,
thanks to the axisymmetry of the background spacetime given by Eq.\ \eqref{BianchiI},
the linear perturbations can actually be decomposed into scalar and vector modes only,
with respect to rotations of the $y-z$ plane,\footnote{Since there is no spherical symmetry in an anisotropic background,
the usual (three-dimensional) scalar-vector-tensor modes cannot develop independently.
However, thanks to one rotational symmetry, the one in the $y-z$ plane, a (two-dimensional) scalar-vector decomposition works well.
For example, the $x$ component of a three-dimensional vector is just a scalar,
and the $y$ and $z$ components can be decomposed into one scalar (gradient) mode and one vector (transverse) mode.
The reason for the absence of (two-dimensional) tensor modes is just that $2\times 2$ symmetric tensors can be written as
two scalar modes and one vector mode only.
In this sense, the ten components of the metric tensor can be decomposed into seven scalar modes and three vector modes generally.
Since three scalar modes and one vector mode can be killed by gauge symmetry, we have four scalar modes and two vector modes in the metric tensor.}
\begin{equation}
 \delta g_{\mu\nu} = \delta g_{\mu\nu}^{\mathrm{scalar}} + \delta g_{\mu\nu}^\mathrm{vector}~.
\end{equation}
The Fourier components of $\delta g_{\mu\nu}^\mathrm{scalar}$ and $\delta g_{\mu\nu}^\mathrm{vector}$ are given by
\begin{widetext}
\begin{equation}
 \delta g^\mathrm{scalar}_{\mu\nu, \vec{k}} =
 \begin{pmatrix}
 - 2 \Phi_{\vec{k}} & *  & *  & 0\\ 
a (\mathrm{i} k_x B_{\vec{k}} - \mathrm{e}^{-4\delta} \frac{k_y}{k} \beta_{1,\vec{k}} )  &  a^2 \mathrm{e}^{-6 \delta} \frac{k_y^2}{k^2} h_{+,\vec{k}}  & * &0 \\
a (\mathrm{i} k_y B_{\vec{k}} + \mathrm{e}^{2\delta}  \frac{k_x}{k} \beta_{1,\vec{k}} )  &  - a^2 \frac{k_x k_y}{k^2} h_{+,\vec{k}}& a^2 \mathrm{e}^{6 \delta}\frac{k_x^2}{k^2} h_{+,\vec{k}} &0 \\
  0 & 0 & 0 & - a^2 \mathrm{e}^{2\delta}  h_{+,\vec{k}} 
 \end{pmatrix}~,
\end{equation}
\end{widetext}
\begin{equation}
 \delta g_{\mu\nu, \vec{k}}^\mathrm{vector}
 =
 \begin{pmatrix}
  0&0 &0 &a \mathrm{e}^{\delta} \beta_{2,\vec{k}} \\
  0&0 &0 &- a^2 \mathrm{e}^{ -4 \delta} \frac{k_y}{k} h_{\times,\vec{k}}  \\
  0&0 &0 &a^2 \mathrm{e}^{ 2 \delta} \frac{k_x}{k} h_{\times,{\vec{k}}} \\
 * &* &0 & 0
 \end{pmatrix}~,
\end{equation}
where $k$ is defined by
\begin{equation}
 k^2 := \mathrm{e}^{4 \delta} k_x^2 + \mathrm{e}^{-2\delta} k_y^2.
\end{equation}
Here we have fixed the arbitrariness of rotations in the $y-z$ plane so that $\vec{k} = (k_x, k_y, 0)$.
We have also fixed the gauge degrees of freedom of perturbations. Our gauge choice corresponds to the spatially flat gauge
[defined by Eqs.\ \eqref{eq:deftensor},\ \eqref{eq:defvector}, and\ \eqref{eq:defscalar}]
in the limit where $\delta \rightarrow 0$ because the perturbed metric reduces to
\begin{equation}
 \lim_{\delta\rightarrow 0}\delta g_{\mu\nu,\vec{k}} = U^{T}(\vec{k}) \Delta(\vec{k}) U(\vec{k})~,
\end{equation}
with
\begin{equation}
 \Delta(\vec{k})=
 \begin{pmatrix}
  -2 \Phi_{\vec{k}} & a \mathrm{i} k B_{\vec{k}} & a \beta_1 & a \beta_2 \\
 * & 0 & 0 & 0 \\
 * & * &  a^2 h_{+} & a^2 h_{\times} \\
 * & * & * &  -a^2 h_{+}
 \end{pmatrix}~.
\end{equation}
The rotation matrix $U(\vec{k})$ is defined by
\begin{equation}
  U(\vec{k}) :=
  \begin{pmatrix}
   1 & 0 & 0 & 0\\
   0 & \frac{k_x}{k} & \frac{k_y}{k} &0 \\
   0 & -\frac{k_y}{k}& \frac{k_x}{k} &0 \\
   0 &0 &0 &1 
  \end{pmatrix}~,
\end{equation}
and it transforms the vector $(0,k_x,k_y,0)$ into $(0,k,0,0)$.

Since the time derivative of $B_{\vec{k}}$ only appears through the following combination in the second-order action,
\begin{equation}
 \varphi_{\vec{k}} := \Phi_{\vec{k}} - \frac{\mathrm{e}^{4 \delta}k_x^2 + \mathrm{e}^{-2 \delta }k_y^2}{3 a H} B_{\vec{k}}~,
\end{equation}
it is useful to regard $\varphi$ as a dynamical variable instead of $\Phi$, analogous to Eq.\ \eqref{varphi=}. 
Thus, we now have six scalar mode perturbations
($\varphi$, $B$, $\beta_1$, $h_{+}$, $\delta \chi_1$, and $\delta \chi_2$),
and two vector mode perturbations ($\beta_2$ and $h_\times$).
By a straightforward calculation,
one can show that the second-order action does not include any time derivatives of $B$,
$\beta_1$, $\delta \chi_1$, $\delta \chi_2$, and $\beta_2$ after integration by parts.
Thus, these variables are nondynamical and the remaining dynamical degrees of freedom are $\varphi$ and $h_{+}$, which are scalar modes, and
$h_{\times}$, which is a vector mode. This result shows that no additional degrees of freedom appear at least in this anisotropic background.
Therefore, our new model is not disturbed by the deformation of the background spacetime given by Eq.\ \eqref{BianchiI}.
This is a crucial difference compared to the model of Sec.\ \ref{sec:models1and2} or $f(R, {\cal G})$ gravity.

\subsection{Recovering Einstein gravity and the addition of matter sources}\label{sec:backgroundmodel3}

At this point, it looks like the action\ \eqref{eq:masteraction} with curvature invariants\ \eqref{eq:I12}
and\ \eqref{eq:I22} can lead to interesting nonsingular cosmological background models
such as an inflationary model and a genesis model that would remain stable against all types of instabilities throughout most of their evolution.
These models start in de Sitter or Minkowski spacetime, respectively, and both end up in Minkowski spacetime.
To be viable structure formation scenarios for the very early universe, as we already pointed out, one needs a reheating mechanism
that would produce radiation and matter in a sufficient amount after the universe has acquired its adiabatic, scale-invariant curvature perturbation
power spectrum. At this point, let us suppose that such a reheating mechanism exists and that it produces matter and radiation in large amounts.
Then, at the background level, one is left with the nonsingular theory described in Sec.~\ref{sec:setupmodel1and2}
with nonzero energy density, and possibly, nonzero pressure as well. For the theory to successfully describe our universe,
it is then necessary to recover the usual Einstein equations, i.e.~the Friedmann equations in our context.

For the inflationary scenario, this is not a problem. Once matter is included, and as $\chi_1$ and $\chi_2$ (and
their time derivatives) go to 0, one notes from Eqs.~\eqref{eq:V1inf} and\ \eqref{eq:V2inf} that $V_1\rightarrow 0$ and $V_2\rightarrow 0$.
Thus, we see that Eqs.~\eqref{eq:Friedmann1} and\ \eqref{eq:Friedmann2} reduce to the Friedmann equations
and can lead to the expected radiation- and matter-dominated era of our universe.

In the context of the genesis scenario, one runs into difficulty though. Once reheating has occurred and matter is included,
one remains in the regime $\chi_2\ll 1$, but $\chi_1\rightarrow\infty$. From Eqs.~\eqref{eq:V1gen}
and\ \eqref{eq:V2gen}, this still implies $V_1\rightarrow 0$ and $V_2\rightarrow 0$. Then, simply at the level of the action\ \eqref{eq:masteraction},
one notices that recovering the Hilbert-Einstein term alone is only possible if $\chi_1I_1\rightarrow 0$,
which is to say that $I_1$ vanishes faster than $\chi_1\rightarrow\infty$.
However, since $I_1\sim\mathcal{O}(R)$, it is not possible to have a nonzero Hilbert-Einstein term while $I_1\rightarrow 0$.
One can also see that, for Eq.~\eqref{eq:Friedmann1} to be valid as $\chi_1\rightarrow\infty$,
the only possibility is that $H$ and $\varepsilon$ vanish faster than $\chi_1\rightarrow\infty$.
Once again, this implies an empty Minkowski spacetime rather than a FLRW spacetime.
Accordingly, it seems impossible, in the context of this genesis scenario,
to have the higher-derivative terms from the curvature invariants vanish, i.e.~to recover the Einstein equations, and
be left with a nonempty Friedmann universe. Therefore, the genesis scenario within this theory remains at the level of a toy model.

It should be noted that nontrivial couplings between matter fields and $\chi_1$ or $\chi_2$ may relax this problem \cite{Easson:1999xw}.
In this case, the bounds on the curvature are, however, weakened because $I_1$ and $I_2$ include matter fields.
Such matter couplings must be subdominant at high energy scales in order to ensure the avoidance of curvature singularity,
but they must be dominant at low energy scales in order to recover Einstein gravity.

\section{Conclusions and discussion}\label{sec:discussion}

In this paper, we revisited the nonsingular cosmologies of Refs.\ \cite{Mukhanov:1991zn,Brandenberger:1993ef},
which implement the limiting curvature hypothesis. We extended the analysis beyond the background cosmology to
include the linear cosmological perturbations and determined the criteria for stability.
This showed that the original models of Refs.\ \cite{Mukhanov:1991zn,Brandenberger:1993ef} appear to have, generically, undesired
additional degrees of freedom leading to Ostrogradski instabilities. These instabilities could be killed with the addition
of the Weyl tensor squared in the curvature-invariant functions given the appropriate coefficient.
Still, by exploring two nonsingular cosmological scenarios in which the limiting curvature hypothesis is realized
(one inflationary and one genesis scenario),
it appeared that the cosmologies inevitably possess either ghost or gradient instabilities through large portions of their
evolution. Furthermore, we showed that the theory could be rewritten as a $f(R,\mathcal{G})$ theory of gravity, which is known
to suffer from instabilities in anisotropic backgrounds.

We then constructed a new curvature-invariant function by taking a specific combination of covariant derivatives of the Ricci scalar.
Given the same inflationary and genesis scenarios at the background level as before, we showed that the new curvature scalar could
lead to stable cosmologies with respect to Ostrogradski ghosts, as well as ghost and gradient instabilities throughout most of their evolution.
Furthermore, the theory does not possess additional new degrees of freedom around anisotropic backgrounds, contrary to
$f(R,\mathcal{G})$ gravity.

In light of constructing a nonsingular theory for the very early universe,
there remain some challenges though. If one starts in a vacuum universe (either de Sitter or Minkowski),
one would need to provide some form of reheating mechanism, possibly via gravitational particle production,
so that the universe can contain matter and radiation after the early epoch. Furthermore, one would
need to assure that the theory reduces to the Einstein limit for gravity fast enough. This appears to be satisfied in our inflationary
scenario, but it remains an issue in the genesis scenario. Finally, given a successful scenario at the background level
and which is stable perturbatively, it would be straightforward to solve the perturbation equations to find the power spectra
of these perturbations and compare with observations to validate the theory.

The analysis performed in this paper opens the window to construct and study other nonsingular cosmologies,
e.g., bouncing cosmologies. As mentioned before, this has been explored in Ref.\ \cite{Chamseddine:2016uef}
(see also Refs.\ \cite{Liu:2017puc,Bodendorfer:2017bjt}).
However, other nonsingular models with limiting curvature\ \cite{Chamseddine:2016ktu,Chamseddine:2016uef}
could also suffer from instabilities as suggested by Ref.\ \cite{Kluson:2017iem}. This might be due to the fact
that Refs.\ \cite{Chamseddine:2016ktu,Chamseddine:2016uef} implement the limiting curvature hypothesis
within a mimetic theory\ \cite{Chamseddine:2013kea,Chamseddine:2014vna,Chamseddine:2016uyr,Sebastiani:2016ras},
whose stability (or instability) does not appear to have been settled yet
(see, e.g., Refs.\ \cite{Barvinsky:2013mea,Chaichian:2014qba,Ramazanov:2016xhp,Achour:2016rkg,Sebastiani:2016ras,Firouzjahi:2017txv}).

Finally, it would be interesting to study how the approach to implement the limiting curvature hypothesis used in this paper
fits in the grand picture of general scalar-tensor theories of gravity (e.g., \cite{Achour:2016rkg,BeyondHorndeski}).
In particular, it would be interesting to find general classes of curvature-invariant functions
in which the limiting curvature hypothesis can be realized and where solutions are stable.

\begin{acknowledgments}
We are grateful to Lavinia Heisenberg and David Langlois for valuable
discussions. D.\,Y.~is supported by the Japan Society for the Promotion of Science (JSPS) Postdoctoral Fellowships for
Research Abroad.  J.\,Q.~wishes to thank the Tokyo Institute of Technology
and RESCEU at the University of Tokyo for kind hospitality while this
work was prepared and acknowledges financial support from the Walter
C. Sumner Memorial Fellowship and from the Vanier Canada Graduate
Scholarship administered by the Natural Sciences and Engineering
Research Council of Canada (NSERC). M.\,Y.~wishes to thank the Institute for
Theoretical Studies (ITS) of the ETH Z\"urich and McGill University for
kind hospitality while this work was initiated and prepared. M.\,Y.~is
supported in part by the JSPS Grant-in-Aid for Scientific Research
Nos.~25287054 and 26610062 and by the MEXT Grant-in-Aid for Scientific
Research on Innovative Areas ``Cosmic Acceleration'' No.~15H05888. R.\,B.~wishes
to thank the ITS of the ETH Z\"urich for a Senior Fellowship
during the 2015/2016 academic year when this project was initiated and
acknowledges financial support from an NSERC Discovery Grant, from the
Canada Research Chair program, from a Simons Foundation fellowship, and
(at the ITS) from Dr.~Max R\"ossler, the Walter Haefner Foundation and
the ETH Z\"urich Foundation.  This work was also partially supported by
the Open Partnership Joint Projects Grant of JSPS.
\end{acknowledgments}

\end{document}